\pdfoutput=1
\documentclass{emulateapj-rtx4}
\usepackage{apjfonts}
\usepackage{amssymb}
\usepackage{natbib}
 \usepackage{color}

\usepackage{graphicx,mathrsfs,amsmath}

\shorttitle{Modeling the distribution of M\lowercase{g} II absorbers around galaxies}
\shortauthors{Bordoloi et al.}

\begin{document}
\title{Modeling the distribution of M\lowercase{g} II absorbers around galaxies using Background Galaxies \& Quasars}
\author{R. Bordoloi\altaffilmark{1}, 
S. J. Lilly\altaffilmark{1},
 G. G. Kacprzak\altaffilmark{2,3}
   \& C. W. Churchill\altaffilmark{4}
 }

\email{rongmonb@phys.ethz.ch}
 
\altaffiltext{1}{Institute for Astronomy, ETH Z\"{u}rich, Wolfgang-Pauli-Strasse 27, 8093, Z\"{u}rich, Switzerland}
\altaffiltext{2}{ Swinburne University of Technology, Victoria 3122, Australia}
\altaffiltext{3}{ Australian Research Council Super Science Fellow}
\altaffiltext{4}{ New Mexico State University, Las Cruces, NM 88003}


\begin{abstract}

We present joint constraints on the distribution of Mg II absorption around high redshift galaxies obtained by combining two orthogonal probes, the integrated Mg II absorption seen in stacked background galaxy spectra and the distribution of parent galaxies of individual strong Mg II systems as seen in the spectra of background quasars. We present a suite of models that can be used to predict, for different two and three-dimensional distributions, how the projected Mg II absorption will depend on a galaxy's apparent inclination, the impact parameter $b$ and the azimuthal angle between the projected vector to the line of sight and the projected minor axis. In general, we find that variations in the absorption strength with azimuthal angles provide much stronger constraints on the intrinsic geometry of the Mg II absorption than the dependence on the inclination of the galaxies.  In addition to the clear azimuthal dependence in the integrated Mg II absorption that we reported earlier in Bordoloi et al (2011), we show that strong equivalent width Mg II absorbers ($W_r(2796) \geq 0.3$ {\AA}) are also asymmetrically distributed in azimuth around their host galaxies: 72\% of  the absorbers in Kacprzak et al (2011), and 100\% of the close-in absorbers within 35 kpc of the center of their host galaxies, are located within $50^{\circ}$ of the host galaxy's projected semi minor axis.  It is shown that either composite models consisting of a simple bipolar component plus a spherical or disk component, or a single highly softened bipolar distribution, can well represent the azimuthal dependencies observed in both the stacked spectrum and quasar absorption  line datasets within 40 kpc.  Simultaneously fitting both data sets, we find that in the composite model the bipolar cone has an opening angle of $\sim 100^{\circ}$ (i.e. confined to within 50$^{\circ}$ of the disk axis) and contains about 2/3 of the total Mg II absorption in the system.  The single softened cone model has an exponential fall off with azimuthal angle with an exponential scale length in opening angle of about 45$^{\circ}$. We conclude that the distribution of Mg II gas at low impact parameters is not the same as that found at high impact parameters. At larger impact parameters beyond 40 kpc, there is evidence for a much more symmetric distribution, significantly different from that closer in to the galaxies. 
\end{abstract}

\section{Introduction}
The absorption line systems observed in the spectra of background sources such as quasars and galaxies are tracers of the otherwise invisible gaseous structures around galaxies. Combining such absorption-line observations with information on their host galaxies, provide a unique and powerful means to understand the baryonic processes in and around galaxies.

The Mg II $\lambda \lambda$ 2796, 2803 doublet is one of the most studied tracers of the gaseous environment around galaxies. These absorption lines are believed to originate in photo ionized gas at temperatures of around T $\sim10^4$  K \citep{Bergeron1986,Charlton2003} and to trace a wide range of neutral hydrogen column densities of $N_{HI} \sim 10^{16} $- $10^{22} cm^{-2}$ \citep{Churchill2000, rigby2002,Rao2006} within a few hundred kpc of their host galaxies.

There have been many detailed studies on the distribution of column densities, the redshift evolution of number densities and the kinematic signatures of Mg II absorbers using quasar absorption line systems (see e.g. \citealt{Lanzetta1987, Sargent1988, Petitjean1990, Steidel1992, Charlton1998, Nestor2005,Prochter2006}). The association of strong Mg II absorption with normal, bright, field galaxies is by now well established (e.g. \citealt{Churchill2005a} and references herein). The Mg II gas around galaxies is traced out to $\sim 100 $ kpc with absorber covering fractions of 50-80$\%$ \citep{Kacprzak2008,Chen2010a,Nielsen2012}. The anti-correlation between absorber equivalent width and impact parameter is also now well established by several studies \citep{Steidel1995, Bouche2006, Chen2010a, Kacprzak2011b, Bordoloi2011a}.

While the association of Mg II absorption with galactic haloes has been secure for some time, the origin and fate of this gas has been less clear.  Possibilities have included outflowing material entrained in star-formation driven winds \citep{Wiener2009,Rubin2010} or inflowing material feeding the disks of galaxies \citep{Rubin2011,Martin2012}.   

The process of large scale outflows driving cold gas out of the galaxy and polluting the circum-galactic medium (CGM) is certainly a complex one. Moreover, the kinematics and the distribution of cold gas entrained in such star formation driven outflows still remain uncertain \citep{Veilleux2005}. The mechanical energy generated by supernovae explosions and cosmic rays etc. create hot over-pressurized bubbles which sweep up the surrounding ambient material into dense shell like structures. These shells accelerate as they expand and when the reach the low density CGM, Rayleigh-Taylor instabilities set in, which leads to fragmentation of these shells in the CGM \citep{Heckman2002,Fujita2009}. 
Although the physical processes involved in such a scenario are likely to be complex, it is likely that cold pockets of gas will be entrained in such outflowing material.  Moreover, new clumps of cold gas might form due to thermal instabilities in the hot gas \citep{Fujita2009}.  Of course, entrained material can also later fall back onto the galaxy due to gravity \citep{Oppenheimer2010}.  Moreover, derived constraints on mass, energetics or momentum of the outflow are hindered with orders of magnitude uncertainties (\citealt{Prochaska2011a} and references therein).
Even idealized outflow models \citep{Fujita2009} are not well constrained. The opening angle of the outflowing gas, the acceleration of the gas as a function of galactocentric radius, optical depth, covering fraction, density, temperature or the spatial extent of such outflowing gas are all quite uncertain. For more complex models, where the Mg II absorption occurs due to combination of outflowing gas with extended galactic disk or inflowing gas, the fractional contribution from each component also remain unconstrained.
\begin{figure*}[!t]
    \includegraphics[height=6.8cm,width=7.5cm]{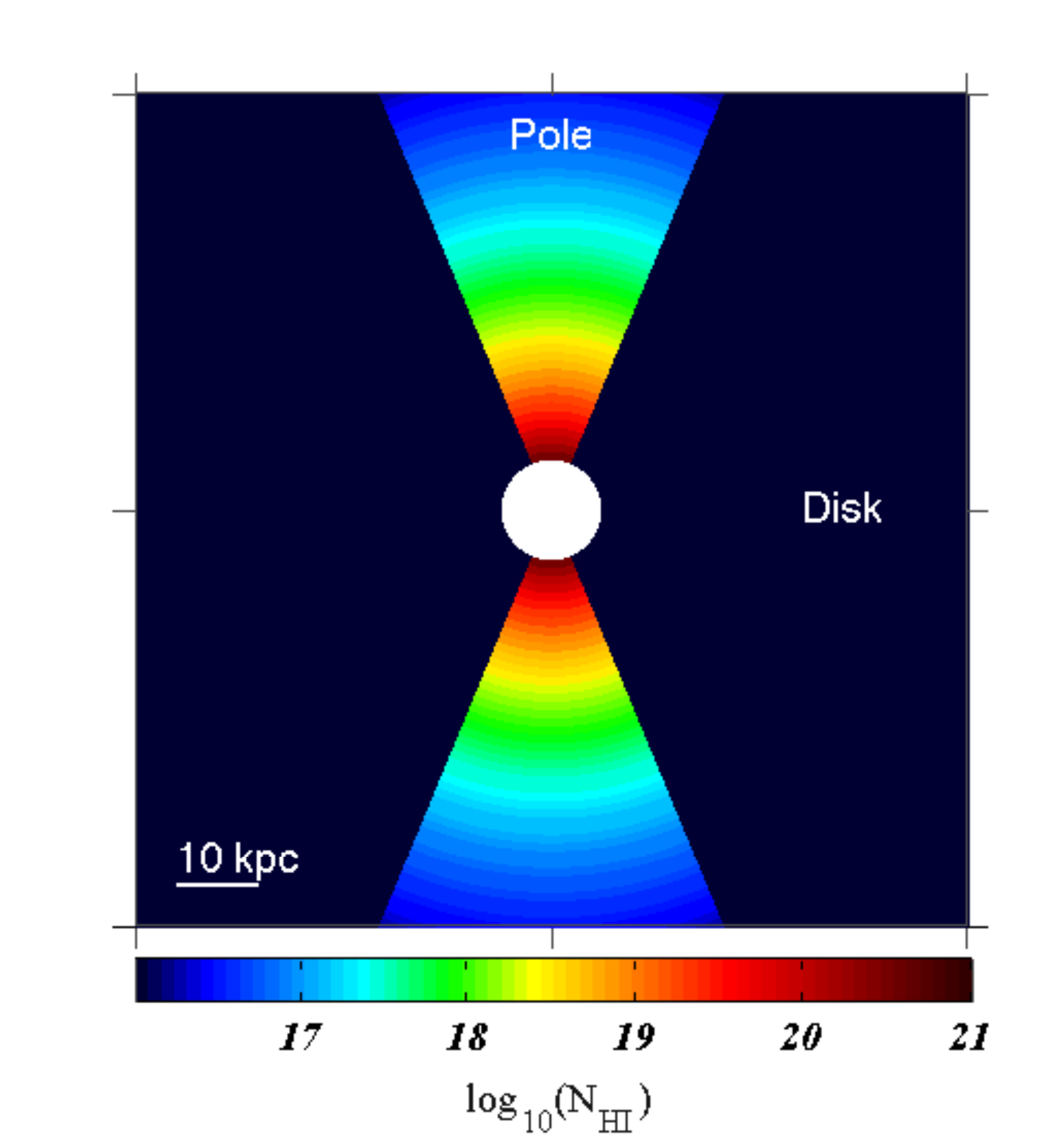}        
    \includegraphics[height=6.8cm,width=7.5cm]{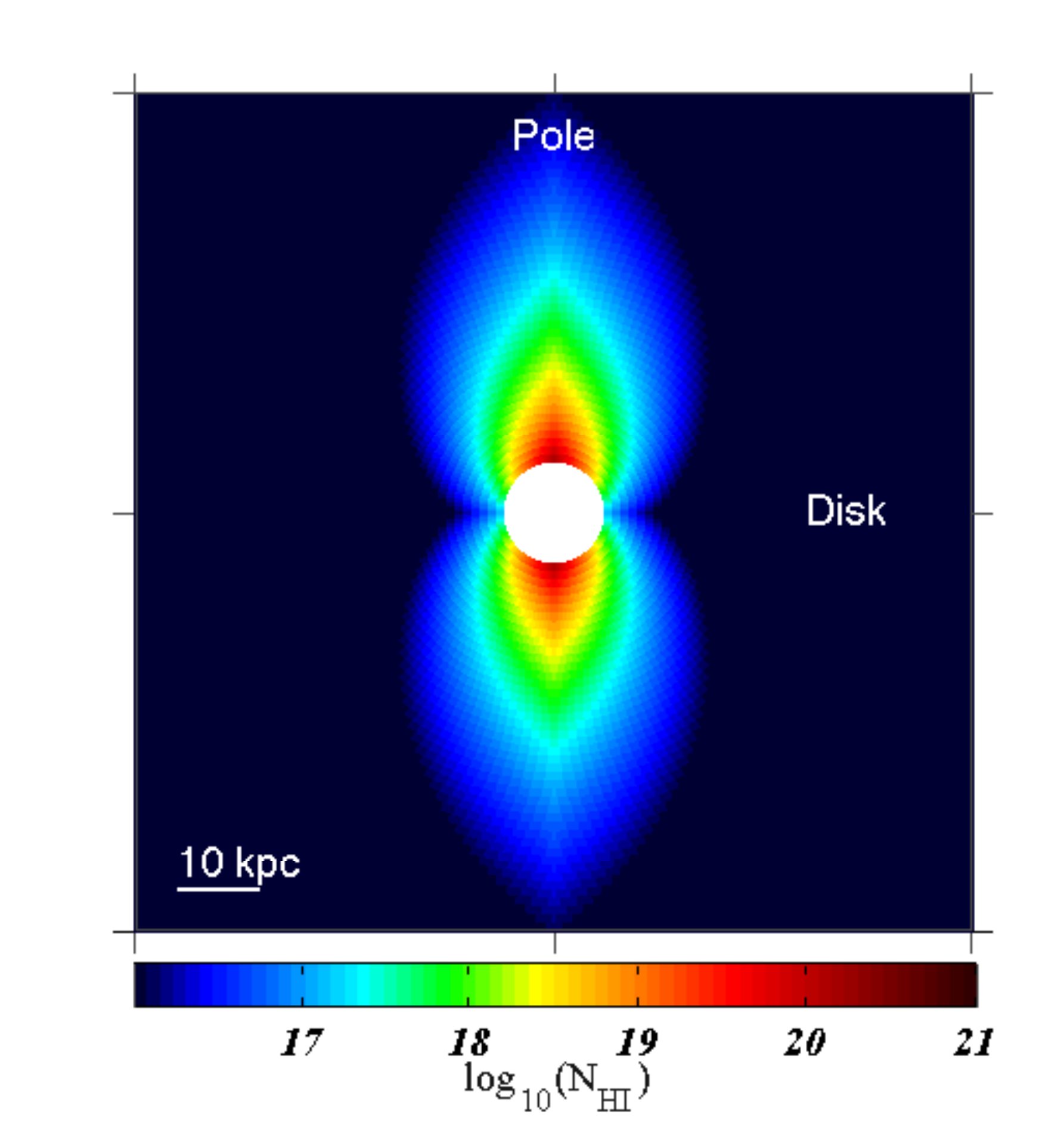}
    \begin{center}
    \includegraphics[height=6.8cm,width=7.5cm]{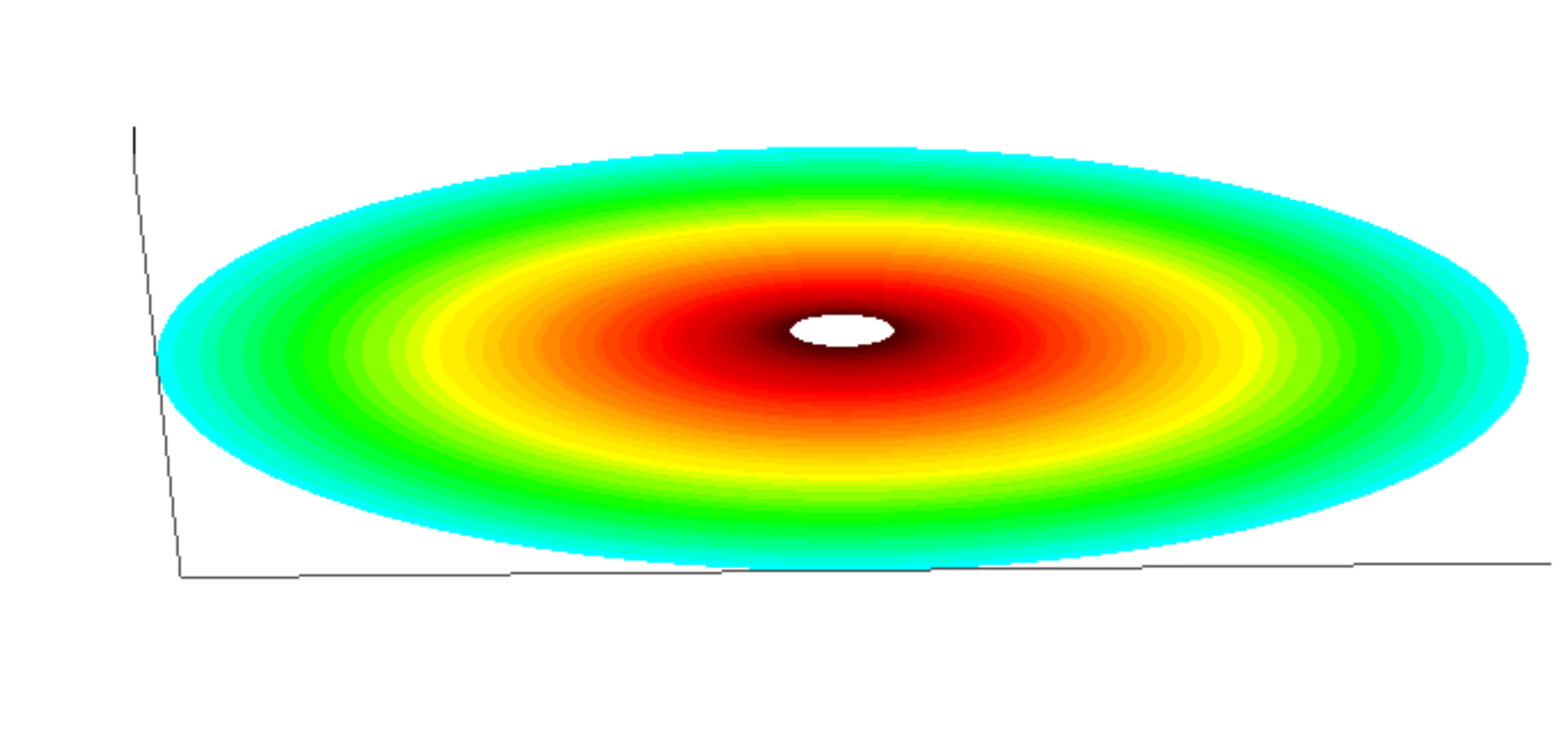}        
       \end{center} 
        \caption{\textit{Top panel:} Cross section of the $N_{HI}$  column density distribution (in the plane of the sky) for the sharp bipolar cone (top left) and for the softened bipolar cone model (top right) both with an opening angle of 45$^{\circ}$ viewed edge on. \textit{Bottom panel:} The projected $N_{HI}$ column density distribution for a 2D disk viewed at an inclination angle. For all three cases, the models have the same projected radial profile, with $R_{cut} = 37.5$ kpc.}
\label{fig1}
\end{figure*}

Recently, several studies have sought to infer the distribution of Mg II absorption around galaxies so as to provide diagnostic information on the origin of the enriched material and to try to constrain possible models.  Using very large numbers of about 5000 background galaxy spectra from zCOSMOS \citep{Lilly2007} to probe the profiles of Mg II absorption around 4000 foreground galaxies at $0.5 < z < 0.9$, \cite{Bordoloi2011a} showed that the strength of the Mg II absorption was much higher for blue star-forming galaxies than for red, more passive, galaxies, at the same stellar mass.  Furthermore, they showed that for inclined foreground disk galaxies, the Mg II absorption is significantly enhanced along the galaxy projected minor axis, at least out to 40-50 kpc from the galaxy.  This result indicated that the Mg II absorption must be primarily originating in a bipolar structure aligned with the disk axis, suggesting that it is part of a bipolar outflow driven by star formation.   

On the other hand, \cite{Kacprzak2011b} found an anti-correlation between Mg II absorber strength and disk inclination $i$, as normalized by the impact parameter $b$, in the sense that the equivalent width of Mg II was anti-correlated with $b/i$. This was interpreted as being due to co-planner geometry of the absorbers.   In parallel \cite{Bouche2011}, using the sample of ten $z \sim 0.1$ absorbers of  \cite{Kacprzak2011a} found a bimodal distribution of azimuthal angles, 60\% of their absorbers being within 30$^{\circ}$  of the galaxy minor axis. They also interpret their result as to support the bipolar wind scenario. \cite{Kacprkaz2012}, using 88 $W_{r}(2796) \geq 0.1 $ {\AA} absorbers and 35 $W_{r}(2796) \leq 0.1 $ {\AA} non-absorbers also reported a bimodal distribution of azimuthal angles: They found that the equivalent width distribution for Mg II absorption along the disk major axis are likely skewed towards the weak absorbers than those along the disk minor axis.~(Also see \cite{Churchill2012}, who find the same result using 51 galaxies with sensitivities to $W_r(2796) \geq 0.003$~{\AA}.

The observational material presented in \cite{Bordoloi2011a} and in the studies of \cite{Kacprzak2011b} and \cite{Bouche2011} are in many ways orthogonal. The integrated approach using stacked galaxy spectra measures the {\it total} Mg II absorption, integrated over all equivalent widths, in some location relative to the host galaxies. This location can be defined over a range of impact parameters $b$ and/or azimuthal angles $\phi$ (which we throughout the paper define to be relative to the projected minor axis). By stacking the spectra to detect the absorption signal in the first place, this total Mg II absorption is by construction measured as an average over some potentially large sample of foreground galaxies that can be constructed as desired in terms of their luminosities, colors, masses, star-formation rates, apparent inclinations and so on.   

In contrast, the quasar absorption line approach involves identifying absorption systems above some individual equivalent width threshold, and then identifying the host galaxies and determining both the location of the quasar  line of sight relative to them, and any properties of the host galaxies of interest (colors, star-formation rates etc.). The distribution of these measurements relative to what would have been expected from a sample of galaxies without the Mg II selection can then give information on the Mg II itself.   A specific example will make the distinction clear.  In the stacked approach, one measures directly the total average Mg II absorption strength as a $f(\phi)$ around galaxies.  In the quasar absorption line approach, one measures the observed distribution of $N(\phi)$ for the hosts of detected Mg II systems and compares this with what would be expected from the general population, i.e. in this case, a flat distribution in $N(\phi)$.  

The main aim of the current paper is to critically examine the information that comes from both the integrated spectra approach and the quasar absorption line approach to see if they present a consistent picture for the distribution of Mg II around galaxies.  Then, since we find that they are indeed consistent, we wish to undertake a joint analysis, using both data sets, to define, as well as possible, the global characteristics of this distribution.   For both purposes, we need a clear understanding of the role of the inclination of a galaxy in determining the Mg II absorption as seen projected onto the plane of the sky.  We therefore construct an extensive suite of models that can be viewed from any angle and with any impact parameter.

The layout of this paper is as follows.  In Section 2 we first discuss the construction of simple geometric models for Mg II around galaxies, including the computation of an integrated equivalent width and the definition of the spatial distributions and geometries of the different models that are considered for the simulations. In Section 3 we ``observe'' these models over the full range of impact parameter $b$, inclination $i$, and azimuthal angle $\phi$, both to identify the most powerful diagnostics for the geometry and to provide predictions for the run of absorption strength with these observable quantities that can be used to compare with the observations. In Section 3.2 we compare these model predictions to the stacked integrated Mg II measurements, including considering straightforward superposition of different components. In Section 4 the model predictions are compared with the quasar absorption line data, yielding similar results. In section 5, we perform a joint analysis of both the datasets to derive our globally preferred model.  Our findings are summarized in Section 6. Throughout this paper, we adopt a $\Lambda$CDM cosmology with $\Omega_{m} = 0.27$, $\Omega_{\Lambda} = 0.73$ and $H_0 = 73.2\;km\; s^{-1}\; Mpc^{-1}$.

\begin{figure}[h!]
\centering
    \includegraphics[height=6.5cm,width=7.5cm]{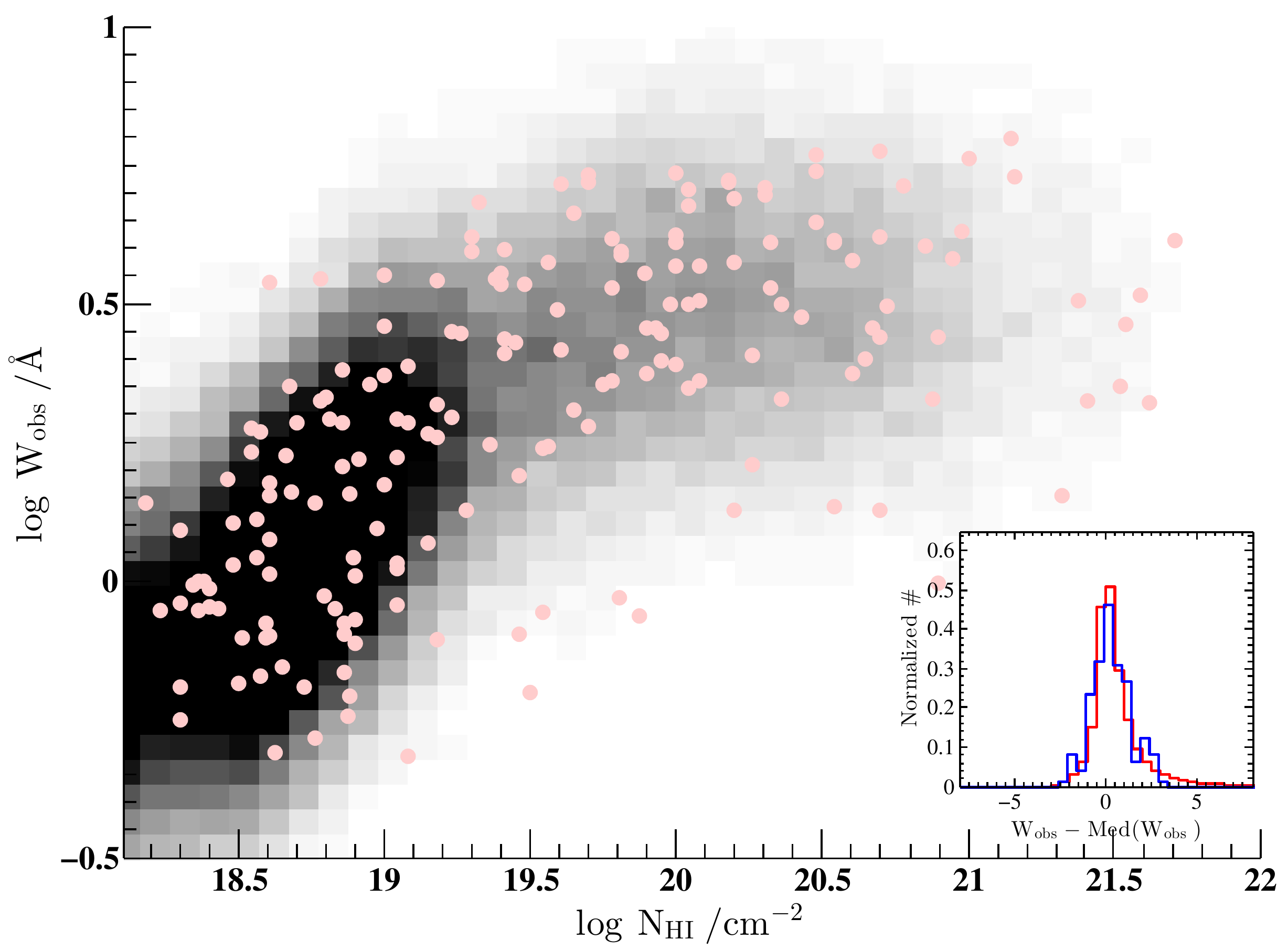}        
    \caption{The relation between $\rm{N_{HI}}$ and $\rm{W_{obs}}$. The pink points show the data from \cite{Rao2006}.The grey background shows the statistical distribution that is used to assign a $\rm{W_{obs}}$ to a modeled $\rm{N_{HI}}$, as described in the text. The inset shows the normalized distribution of ($\rm{W_{obs} -Median (W_{obs})}$), from the \cite{Rao2006} data (blue histogram) and the models (red histogram).  }
\label{fig2}
\end{figure}

\begin{figure}[h!]
\centering
    \includegraphics[height=6.5cm,width=7.5cm]{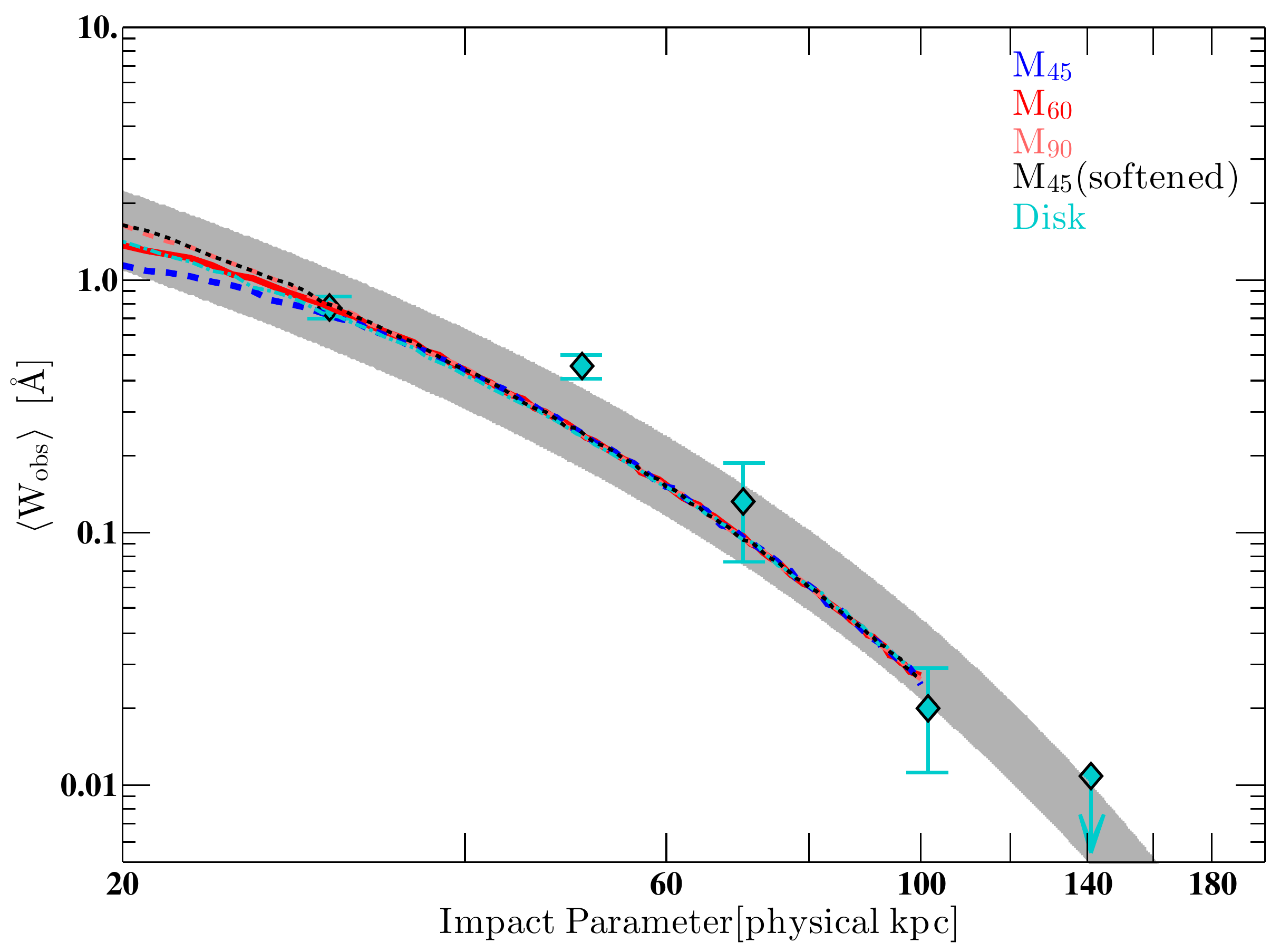}        
    \caption{The azimuthally averaged radial profile for five models studied in this paper (dashed and dotted lines). For all the models the overall normalization ($\alpha$) is so chosen that $\langle W_{obs} \rangle \simeq 0.7$ {\AA} within $20\leq b \leq 50$ kpc. The data points are from \cite{Bordoloi2011a}. The gray shaded region shows the power law with an exponential cut-off fit to the data given in equation 2.}
\label{fig3}
\end{figure}



\begin{figure*}[!htb]
   \includegraphics[height=6.5cm,width=9.cm]{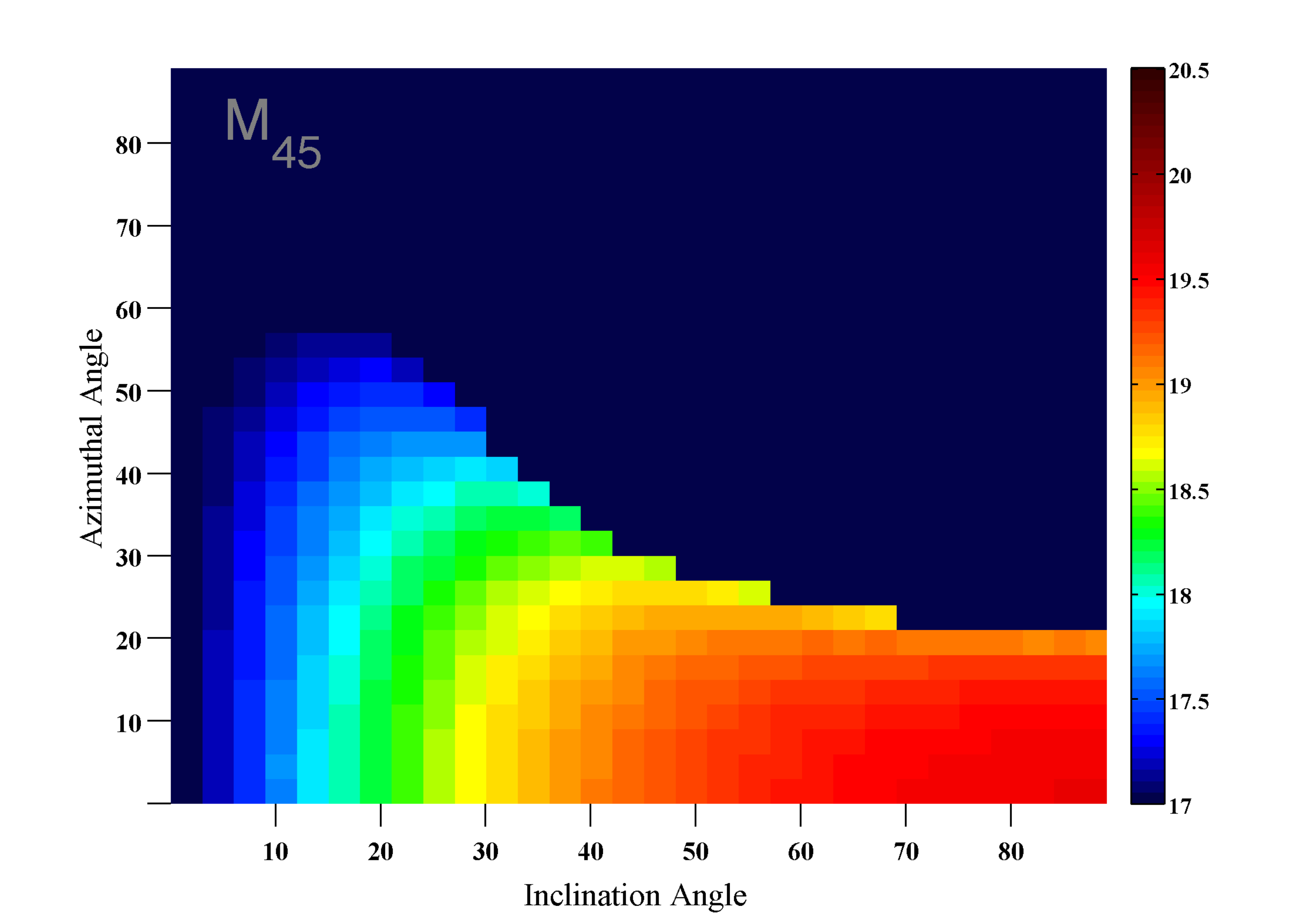}
   \includegraphics[height=6.5cm,width=9.cm]{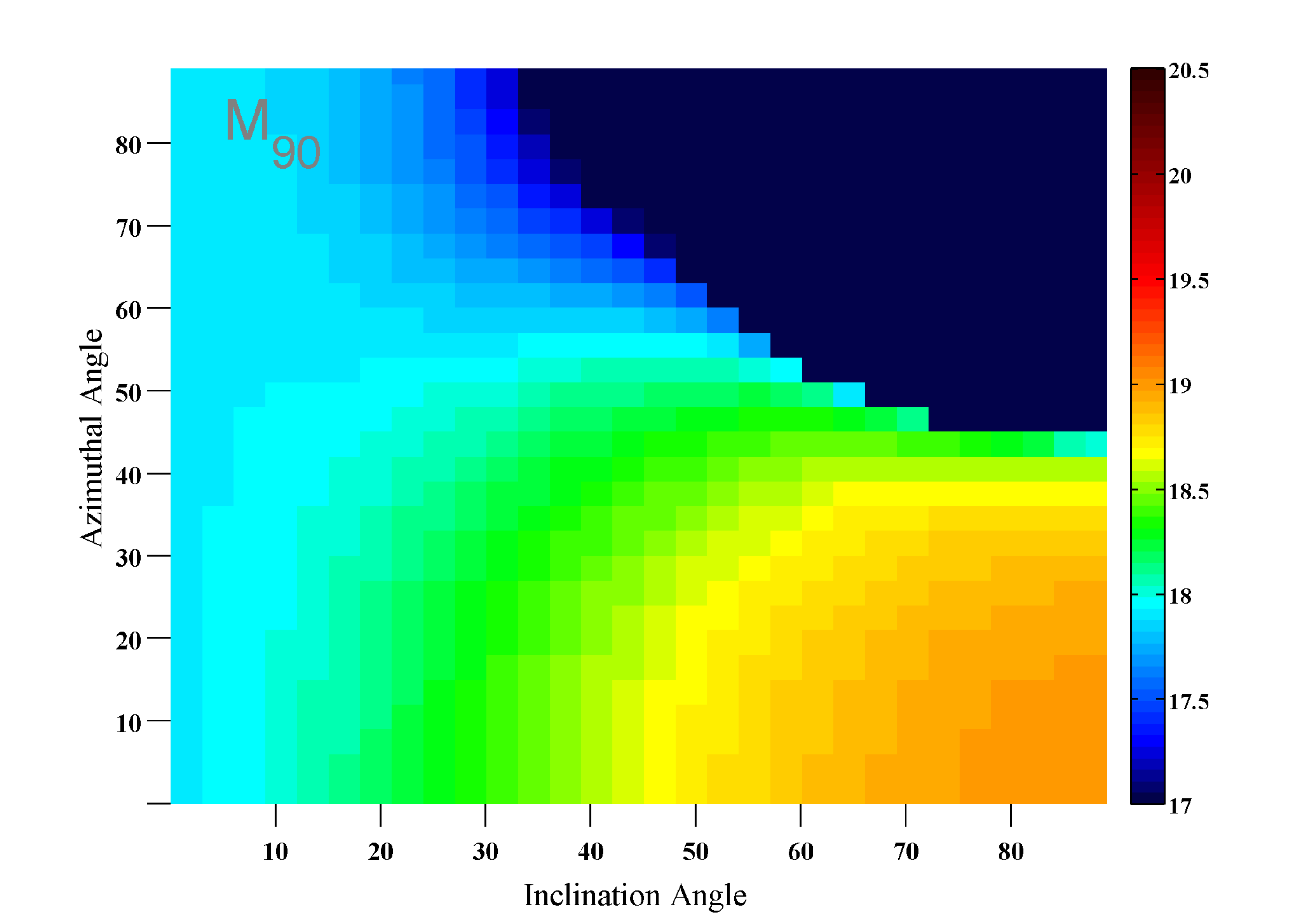}
   \includegraphics[height=6.5cm,width=9.cm]{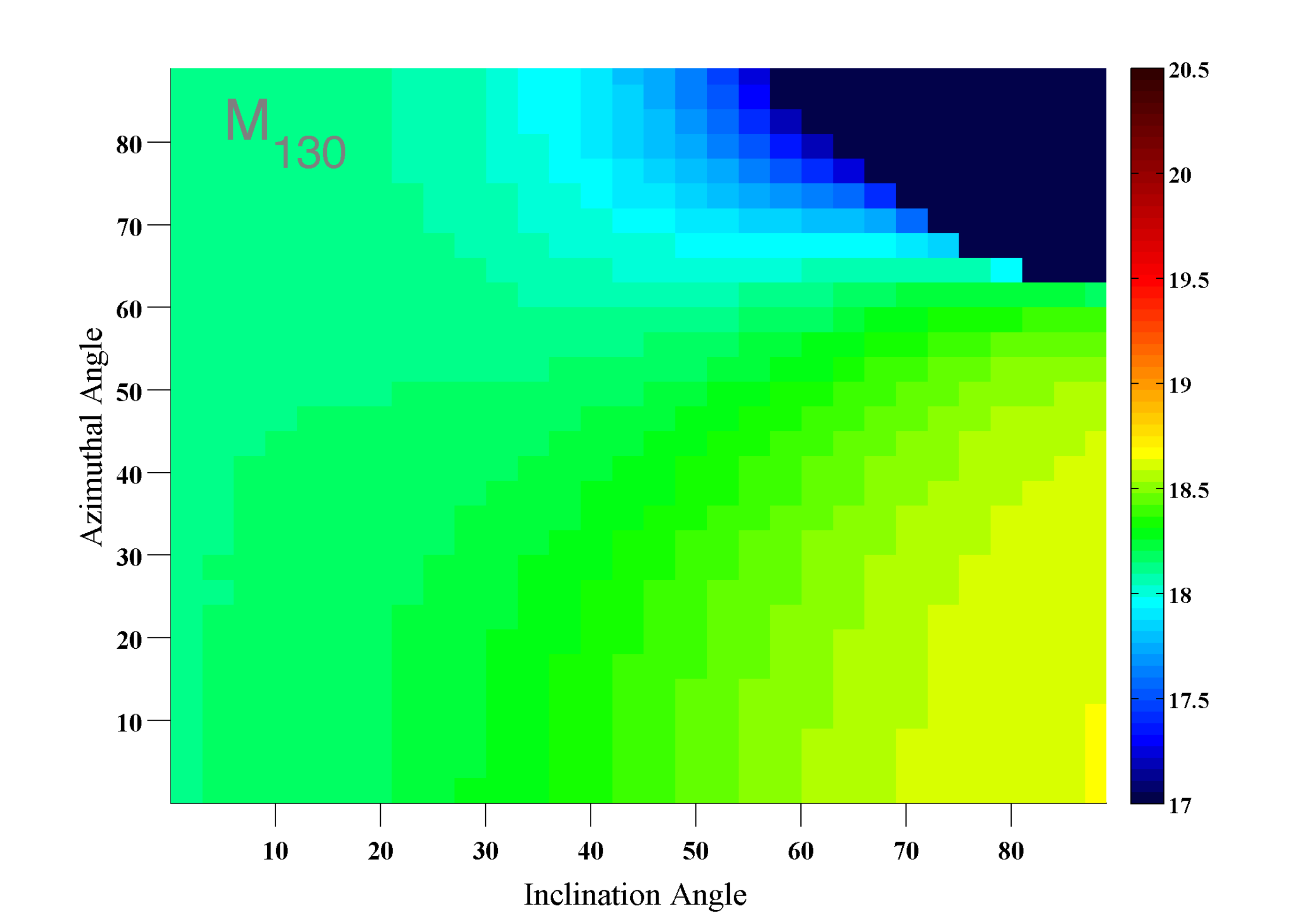}
   \includegraphics[height=6.5cm,width=9.cm]{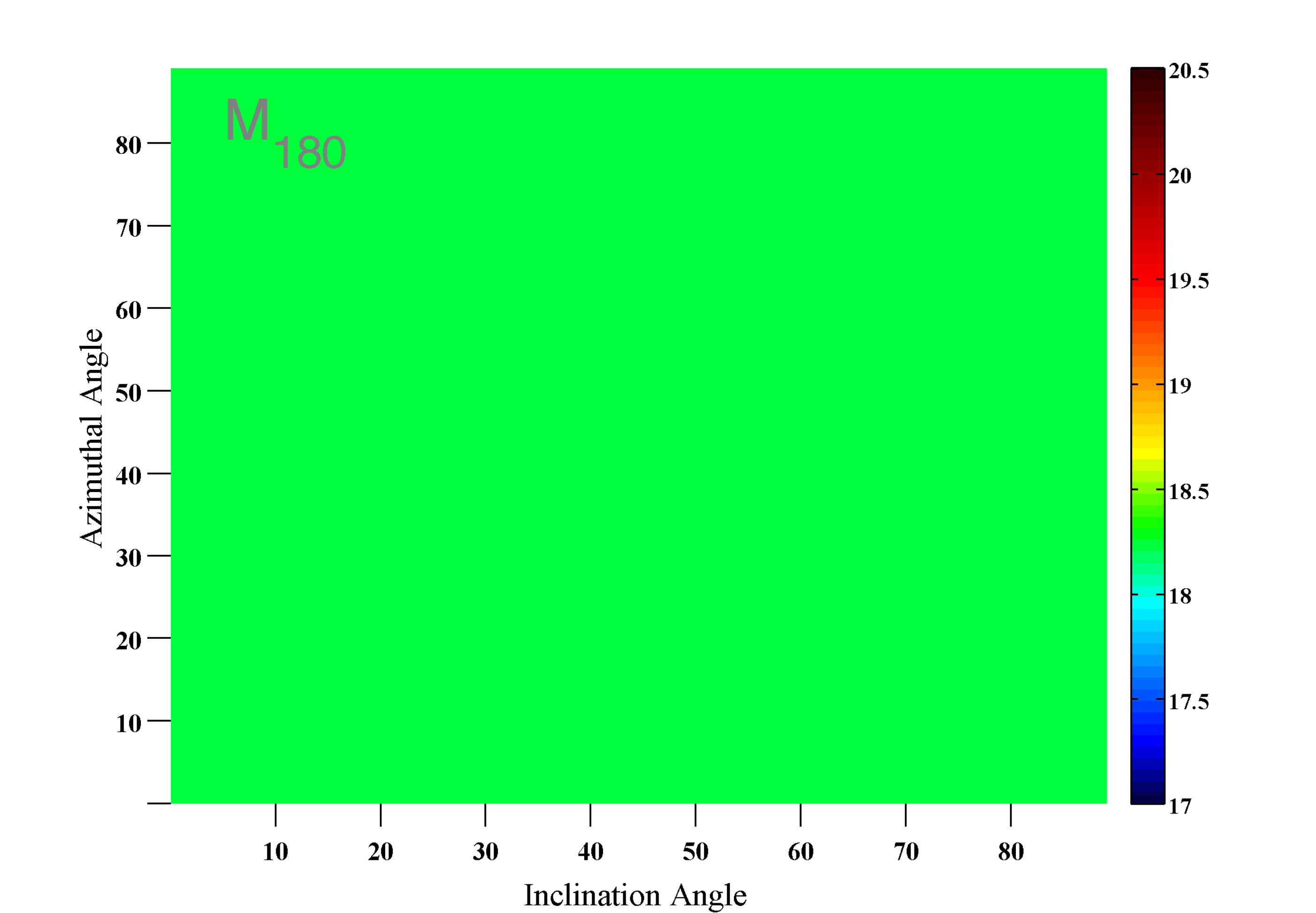}
   \includegraphics[height=6.5cm,width=9cm]{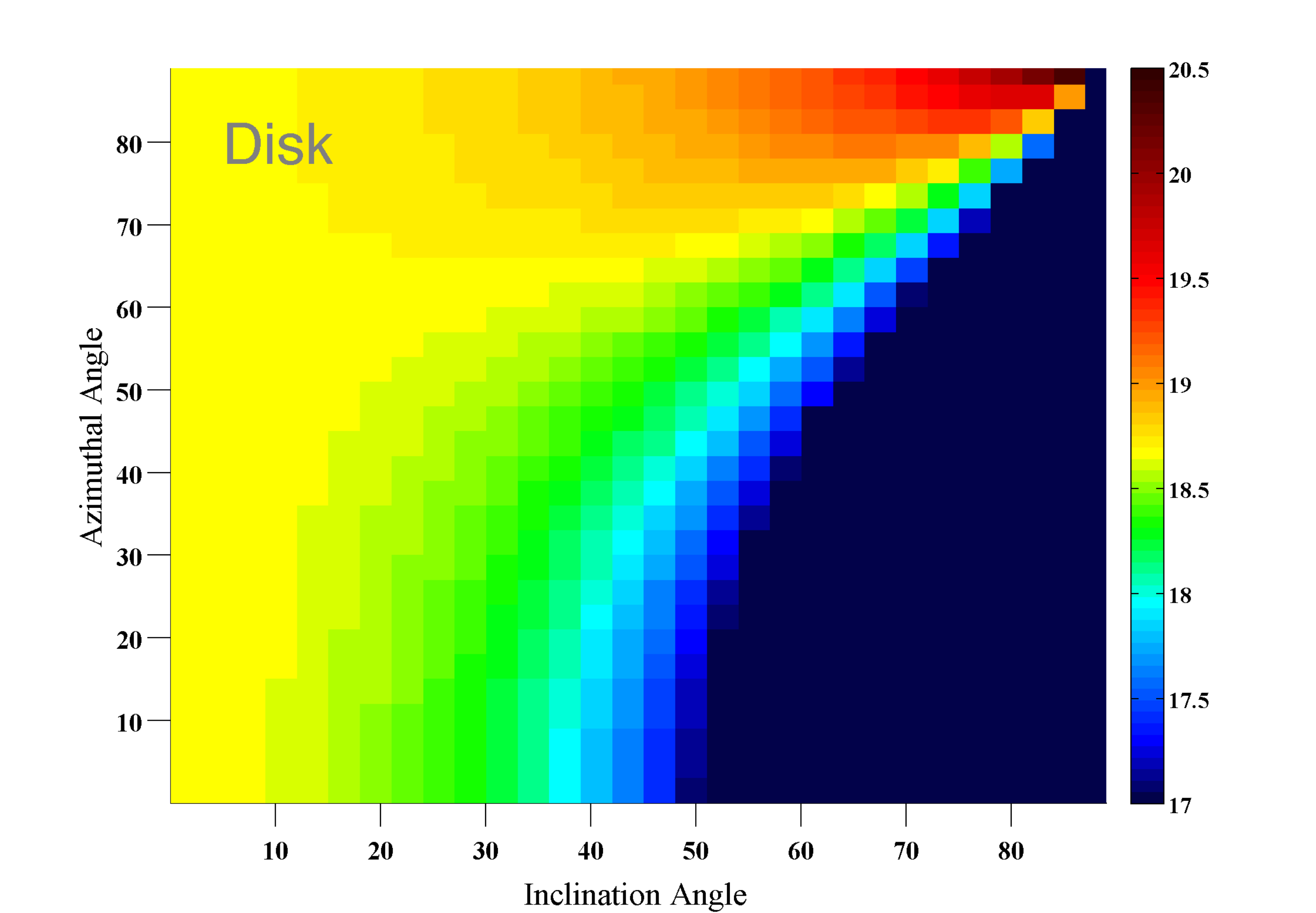}
   \includegraphics[height=6.5cm,width=8.3cm]{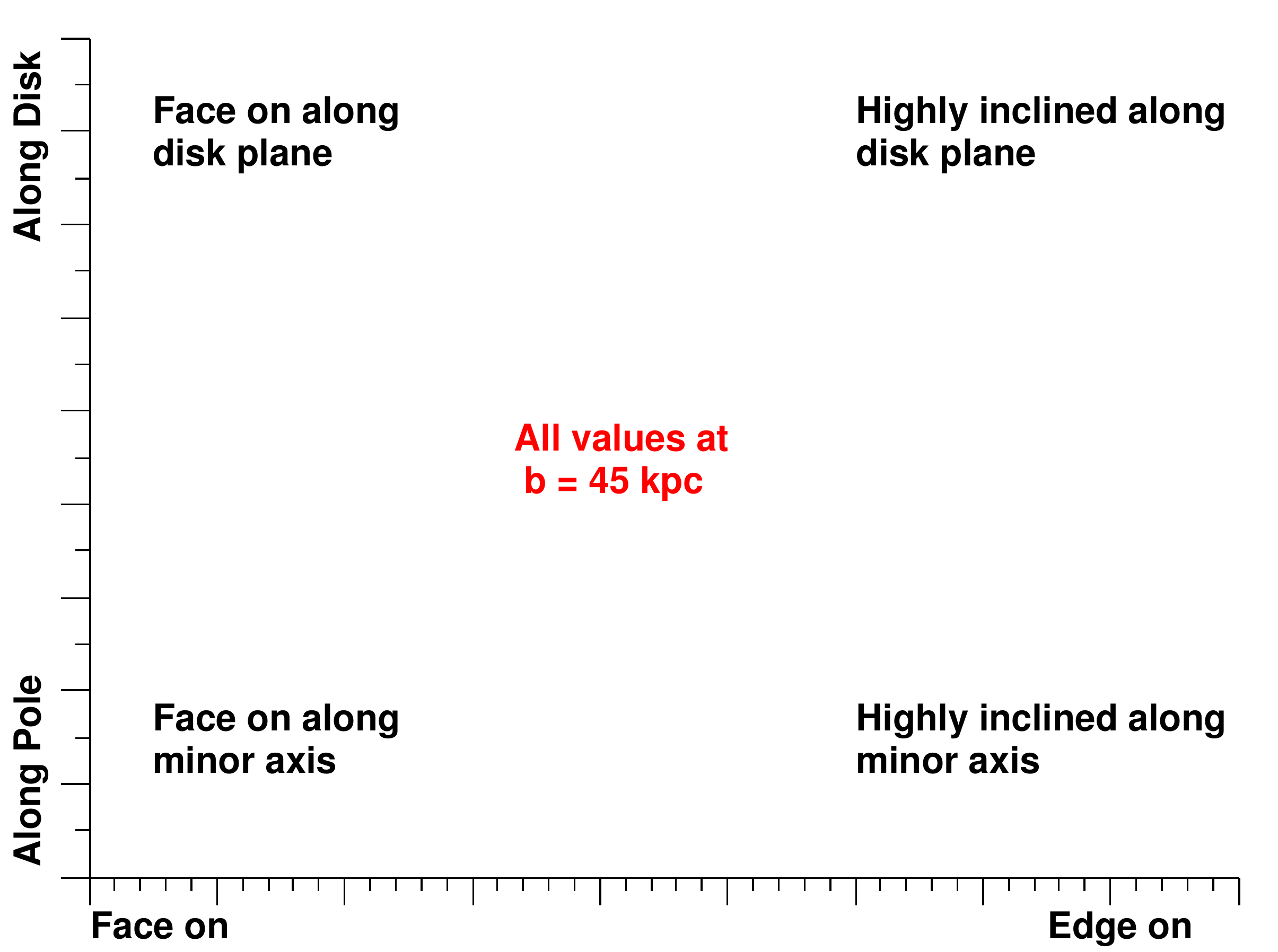}
         \caption{ The $\log_{10} \langle N_{HI}(b) \rangle$ column density distributions for different models as a function of inclination and azimuthal angle, for $b =$ 45 kpc. The models shown here are respectively from top left (i) cone with opening angles ($\chi$) 45$^{\circ}$, 90$^{\circ}$, 130$^{\circ}$, (ii) sphere and (iii) Disk. Different 3-d absorption geometries show clear trends with inclination and azimuthal angles. The last panel shows the orientation of the absorbers. }
\label{fig4}
\end{figure*}

\section{The Model}

We will assume throughout the paper that the light from the galaxy is in a disk and that the ellipticity of this light indicates the inclination of the disk.  We describe this orientation by the direction of the pole of this disk, which lies in the direction $\vec{P}$. 
We then create a set of galaxy models that are each based on a particular 3-dimensional distribution for the absorbing gas that is given in terms of a density $n(\vec R)$ which is the number density of the neutral hydrogen atoms (number of atom per $cm^{-3}$) in the galaxy at position $\vec R$. The angle subtended by $\vec R$ with respect to the polar axis $\vec{P}$ is defined to be $\psi$, i.e. $\cos \psi = \vec{R}\centerdot \vec{P}$. We consider a spherically symmetric distribution, a bipolar distribution and a two-dimensional disk. In each case the density distribution is assumed to be rotationally symmetric about the pole.  

The galaxy is then rotated randomly through $10^6$ orientations, i.e. so that $\vec{P}$ is uniformly distributed on the unit sphere. The expected Mg II absorption is then calculated by integrating through a given line of sight that passes at a projected separation, i.e. impact parameter, $b$ from the centre of the galaxy. The observed column density $N_{HI}(b)$ at impact parameter $b$ is given as

\begin{equation}
N_{HI} (b) = \int \limits_{L\, O\, S} n (\vec{R}) \;dl \;.
\label{eqn1}
\end{equation} 
 
Each of the $10^6$ random orientations is associated with a particular inclination for the observer, where the inclination angle $i$ is defined so that $i= \; 0^{\circ}$ corresponds to a face-on galaxy and  $i= \; 90^{\circ}$ corresponds to an edge-on galaxy whose pole lies in the plane of the sky.  The inclination angle $i$ has a probability distribution $\rm{P(i)\,di \sim \sin(i) \,di}$, i.e. more galaxies are edge-on than face-on.   For simplicity, the lines of sight are chosen to all lie along a particular line in the sky extending from the center of the galaxy, so that each orientation also corresponds to a projected azimuthal angle $\phi$ between the projected pole of the galaxy, i.e. the observed minor axis, and the arbitrary vector in the plane of the sky along which the lines of sight have been calculated.  We use $\phi \;=\;0^{\circ}$ for a sight line that is located along the minor axis, i.e. above the pole of the galaxy, and $\phi \;=\;90^{\circ}$ corresponds to a sightline aligned along the major axis.
This azimuthal angle ($\phi$) will be uniformly distributed between $0^{\circ}\;\leq\;\phi\;\leq\;90^{\circ}$. Throughout this paper, kinematics of the Mg II absorbers are not taken in to account. We quote the total equivalent width of the absorption absorption line system and assume no knowledge of their kinematic velocity spread. All the models will be normalized to the radial profile of blue galaxies from \cite{Bordoloi2011a} (described in the next section), hence the models studied here will represent blue host galaxies within the stellar mass range of $\sim 10^{9.5} M_{\odot}$ to $ 10^{10} M_{\odot}$. Figure \ref{fig1} shows the cross section of three models used in this study which will be described in the next section. 

\subsection{Calculation of $W_{obs}$ }

As described above, the models are constructed using neutral hydrogen number densities, $n(HI)$ and column densities $N_{HI}$.  For a given line of sight, we use the observed distributions of $N_{HI}$ column densities and  Mg II equivalent widths from the study of \cite{Rao2006} to assign an observed Mg II equivalent width ($W_{obs}$) for each line of sight.

In the models, for a particular line of sight with some neutral hydrogen column density $N_{HI}$, we assign a $W_{obs}$ by sampling the \cite{Rao2006} dataset (as shown by the pink points in figure \ref{fig2}).  For each line of sight, we randomly draw a $\log W_{obs}$ from a Gaussian distribution of width 0.2 dex centered on the median $\log W_{obs}$ of the \cite{Rao2006} data. The 0.2 dex spread of the Gaussian distribution is chosen to account for the spread in the data. The inset of figure \ref{fig2} shows the normalized distribution of ($\rm{W_{obs} -Median (W_{obs})})$ of the \cite{Rao2006} data (blue histogram) and the model (red histogram) respectively. The 0.2 dex spread well matches the observed data and a two sample KS test can not rule out the null hypothesis that these two normalized distribution were drawn from the same parent population at 10\% significance. The median $\log W_{obs}$ were obtained by using a one dimensional running median filter with a box size of 19 nearest neighbors along the $\log N_{HI}$ axis. This treatment reflects the general behavior of the Mg II equivalent width distribution and blending of kinematic components in this equivalent width regime (e.g., \citealt{Churchillvogt2003}). The statistical distribution of  $W_{obs}$($N_{HI}$) is shown in figure \ref{fig2} as the gray background. It should be noted that the $W_{obs}$ values quoted here are summed over both the components of the Mg II absorption doublet ($W_{obs} = W_{2796} + W_{2803}$).  For simplicity we assume throughout a doublet ratio of 1:1, i.e. the $W_r(2796)$ component will have half the value of $W_{obs}$.

We next construct a variety of three dimensional distributions of the $n(HI)$.   As detailed below, the radial profiles of these are all chosen to produce a projected radial profile of the same shape as was measured for blue galaxies in \cite{Bordoloi2011a}. This observed radial profile can be well represented by a $\rm{b^{-1}}$ power law with an exponential cutoff given by

 \begin{equation}
 \rm{ \langle W_{obs} \rangle(b) = 1.69 \pm 0.19 \; \left(\frac{b}{35 \; kpc}\right)^{-1} \; \exp\left(-\frac{b}{35 \; kpc}\right)  { \textup{\AA}}}.
 \end{equation}
This is shown as the grey shaded region in figure \ref{fig3}.  This implies a radial fall off in density of $n \propto R^{-2}$ for 3-d model distributions and of $n \propto R^{-1}$ for 2-d distributions, both with suitable exponential cut-offs.  

\subsection{3-D bi-cone and sphere models}

The first model is set up by assuming that absorbing gas is distributed with a conical geometry, aligned along the disk rotation axis. Within the galaxy, the angle with respect to the polar axis is defined to be $\psi$, i.e. $\cos \psi = \vec{R} \centerdot  \vec{P}$.  Within this cone, the number density is given by the galactocentric radius, $R$, and to be zero outside the cone. 

 \begin{equation}
n(R)  =
\begin{cases}
  \alpha \; \left(\frac{R}{R_{cut}}\right)^{\beta} \; \exp\left(-\frac{R}{R_{cut}}\right), & \mbox{if } \psi \leq  \chi /2\\ \\
 0, & \mbox{otherwise .}
\end{cases}
\label{eqn2}
\end{equation}

Here $\chi$ is the opening angle of the bipolar cone. The generic models defined by equation \ref{eqn2} are hereafter referred to as \emph{sharp bipolar cone} models. These models represent the simple model of bi-conical outflow coming out of a galaxy.  We create a suite of models with opening angles 
$\chi \in [0^{\circ}, 180^{\circ}]$.  The model with $\chi = 180^{\circ}$ is therefore a full sphere.

Taking $\beta = -2.0$ to match the average projected radial profile in \cite{Bordoloi2011a}(dashed and dotted lines in figure \ref{fig3}). This is obviously appropriate for a mass-conserving outflow of constant velocity. The default cut-off radius in the exponential ($R_{cut}$) is set at 37.5 kpc, also to match the average radial profile. 

The overall normalisation of the $N_{HI}$ column density in this and all subsequent models is chosen by setting the value of $\alpha$ so that the average $\langle W_{obs} \rangle \simeq 0.7$ {\AA} for $20 \leq b \leq 50$ kpc, i.e. to match the observed average radial profile in figure \ref{fig3}.  Because the conical models all have the same radial dependence, there is a simple relation between the adopted $\alpha$ and the opening angle $\chi$.  

Top left panel of figure \ref{fig1} shows a cross section of the $N_{HI}$ column density distribution of the sharp bipolar cone model with $\chi = 45^{\circ}$ viewed edge on with $i = 0$.  For the remainder of the paper we refer to sharp bipolar cone models with a subscript indicating the opening angle $\chi$, i.e. M$_{30}$ has an opening angle of $30^{\circ}$ and M$_{180}$ is the sphere.  

The second set of bipolar cone models are defined to have the same radial fall-off as previously described, but now with a softened exponential falloff in $\psi$, i.e. 

\begin{equation} 
n(R,\psi)  = \alpha \; \left(\frac{R}{R_{cut}}\right)^{\beta} \; \exp \left(-\frac{R}{R_{cut}}\right) \; \exp\left(-\frac{2 \psi}{\chi}\right)
\label{eqn3}
\end{equation}

The generic set of models defined by equation \ref{eqn3} are hereafter referred to as \emph{softened bipolar cone} models. Again as before, we set $R_{cut} = 37.5$ kpc and $\beta = -2.0$. These models are, as before, normalized to $\langle W_{obs} \rangle \simeq 0.7$ {\AA} for $20 \leq b \leq 50$ kpc, by adjusting the value of $\alpha$. This model has the feature of a non-zero density down to the plane of the galaxy and might represents a situation in which material also flows  back onto the galaxy along the disk plane. The top right panel of figure \ref{fig1} shows a cross section of the $N_{HI}$ column density distribution in such a model viewed edge on with $\chi = 45^{\circ}$. $W_{obs}(b)$. We refer to these models as M$_{\chi}$(softened).

In figure \ref{fig3}, we plot the averaged radial profiles for $M_{30}, M_{45}, M_{90}$ and $M_{45}$(softened) models used in this study (averaged over both inclination and azimuth) compared with the measured radial profile taken from \citep{Bordoloi2011a} (blue data points). Because of the way these models are constructed, all the models have similar average radial profile in $\langle W_{obs} \rangle(b)$. Not surprisingly, the radial profile alone gives no information on the $\psi$ distribution of the absorbing material. The grey shaded region shows the 3$\sigma$ confidence intervals of the $\rm{b^{-1}}$ power law fit to the data given by equation 3. 

\subsection{2-D disk model}

We also define a 2D model consisting of an infinitesimally thin disk, characterized as 

\begin{equation} 
N_{perp}(R)  = \alpha \; \left(\frac{R}{R_{cut}}\right)^{-1} \; \exp\left(-\frac{R}{R_{cut}}\right) \;\;\;\; \rm{;\;along\;\;the\;\;disk}
\label{eqn4}
\end{equation}

where $N_{perp}$ is column density perpendicular to the disk. The cut-off radius ($R_{cut} = 37.5$ kpc) is identical to that of the 3-D models, but because of the reduced dimensionality, the exponent of the power-law is reduced by one. The normalization term $\alpha$ is again chosen similar to that done in the case of the 3-D models. For the 2-D disk case, the observed column density $N(b)$ can be measured directly without integrating along the line of sight but with a correction for non face-on disks because of the increase path through the disk. The radial position where a line of sight intersects the plane of the disk is given in terms of $b$, $i$ and $\phi$ and so, accounting for the longer path length through a inclined disk, we get
\begin{equation}
N(b) = N_{perp}\left(b \left[\sin^2 \phi  + \left(\frac{\cos \phi}{\cos i}\right)^2\right]^{1/2} \right) \times \sec i
\label{eqn5}
\end{equation}

$N(b)$ is converted to $W_{obs}(b)$ as discussed in section 2.1. The model is normalized to have $\langle W_{obs} \rangle \simeq 0.7 \AA$ within $20 \; \rm{kpc} \leq b \leq 50\; \rm{kpc}$ by setting the value of $\alpha$. The bottom panel of figure \ref{fig1} shows the {\it projected} $N_{HI}$ column density distribution of this model viewed at an inclined angle. In figure \ref{fig3} we over plot the average radial profile $W_{obs}(b)$ for the disk model, which is again very similar to that of the 3-D models.  As noted above, the average radial profile alone contains no information to distinguish between 2D and 3D geometries. 

\subsection{Hybrid models}

In this section we define two further, hybrid, models by combining the 3-D and 2-D models described in the previous sections. The first  Hyb1 ($M_{\chi}$ + sphere) model is constructed by linear combination of a sharp bipolar cone model with opening angle $\chi^{\circ}$ and the sphere model. The relative strength of the two components is conveniently parametrized as given below,

\begin{equation}
\eta = \frac{\rho_{cone}}{\rho_{sphere}}
\label{density_ratio}
\end{equation}
 
where $\rho_{cone}$ gives the number density of neutral hydrogen atoms in the cone region and $\rho_{sphere}$ gives the number density of atoms in the rest of the model. Effectively $\eta$ gives the over density of neutral hydrogen atoms in the cone region with respect to the rest of the model. $\eta$ is a free parameter, which can be varied to make the hybrid models have a wide range of density contrast between the cone region and the rest of the model. If $\eta >>$ 1, the cone region dominates the final contribution to the hybrid model. When $\eta=1$, the number density of atoms in the cone region and the rest of the model is the same and hence irrespective of the opening angle, the hybrid model has a uniform number density distribution of a spherical model. If $\eta <$ 1, the number density of atoms in the cone region is less than that in the rest of the model. When $\eta < 1$, the over density of neutral hydrogen atoms will be aligned along the major axis of the disk. As the opening angle increases, the hybrid model approaches a spherical geometry. The hybrid model is normalized to have the same $W_{obs}(b)$ radial profile as shown in figure \ref{fig3}.

The second hybrid model (Hyb2) is constructed also as a combination of  a sharp bipolar cone with opening angle $\chi$ and the disk model. The two components of the hybrid model are weighted as defined below

\begin{equation}
W_{hyb} = \frac{f  \;W_{cone} + W_{disk}}{ 1+ f}
\label{whyb}
\end{equation}

Where $W_{hyb}$ is the contribution to the final hybrid model, $W_{cone}$ is the sharp bipolar cone model with a given opening angle and $W_{disk}$ is the disk model. $f$ is a free parameter which controls the fractional contribution of each of the cone or the disk component. This hybrid model also has the same radial profile as in figure \ref{fig3}.

Physically, the two hybrid models could represent a scenario where the absorbers originate due to a combination of a bipolar outflow component combined with either a spherical halo component (Hyb1) or an extended disk component (Hyb2). 

\section{Observable quantities}

Given the radial profile $\langle W_{obs} \rangle(b)$ for each of the $10^6$ orientations, each with their unique $(i,\phi)$ we average these to yield observable quantities including (1) radial profiles $\langle W_{obs} \rangle (b)$, i.e. averaged over $i$ \& $\phi$, (2) $W_{obs}(i, \phi)$ at a given b, (3) the inclination profile : $\langle W_{obs} \rangle (i)$ for a given $b$, averaged over $\phi$ and (4) azimuthal profiles: $\langle W_{obs} \rangle (\phi)$, at given $b$ and averaged over $i$. This exercise is then repeated for the different geometrical models constructed as described below.

In this section we describe the basic observable quantities for the different model geometries and diagnose which observables can most efficiently discriminate between them.


\begin{figure*}[!ht]
\includegraphics[height=9cm,width=9cm]{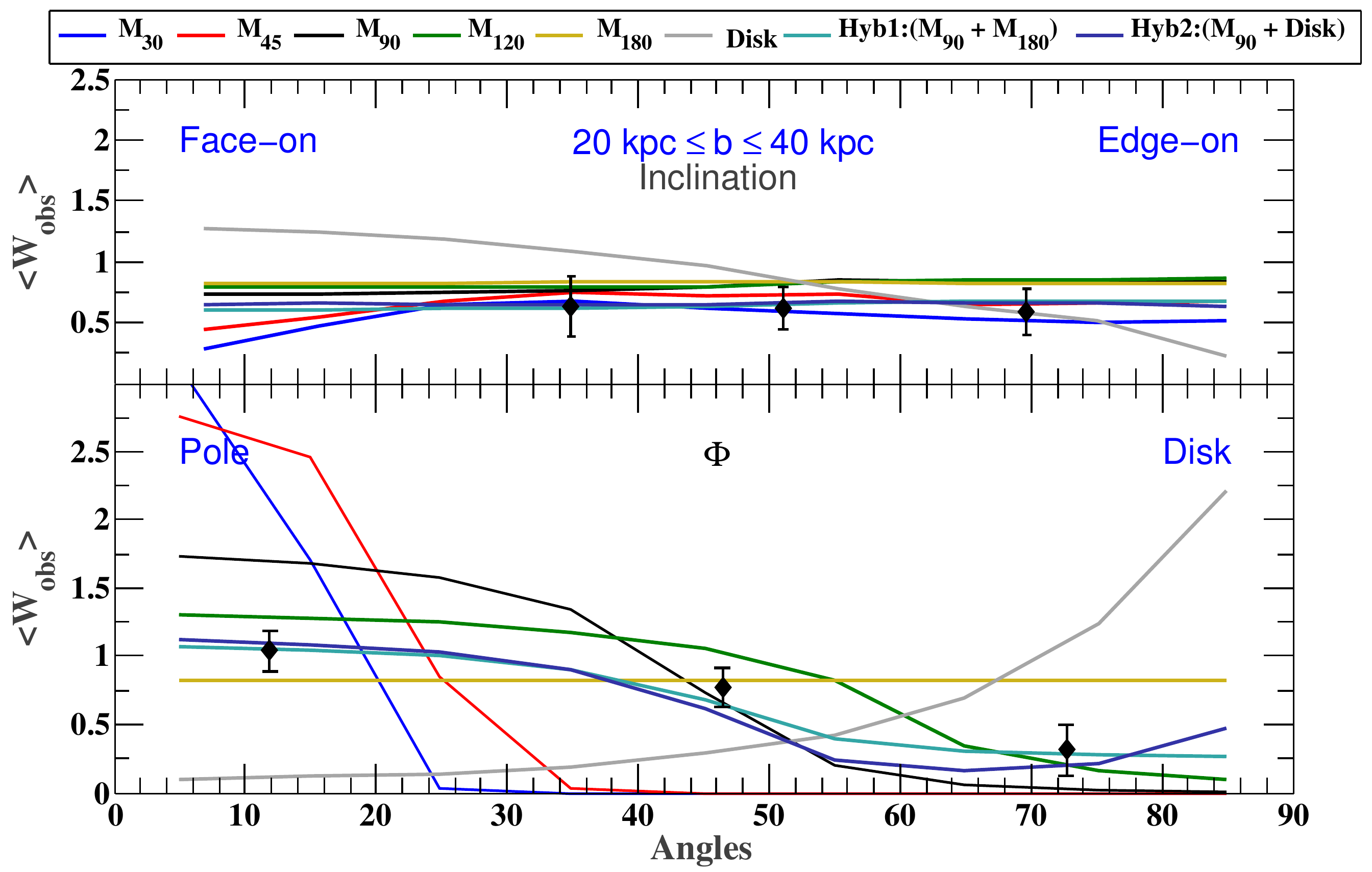}        
\includegraphics[height=9cm,width=9cm]{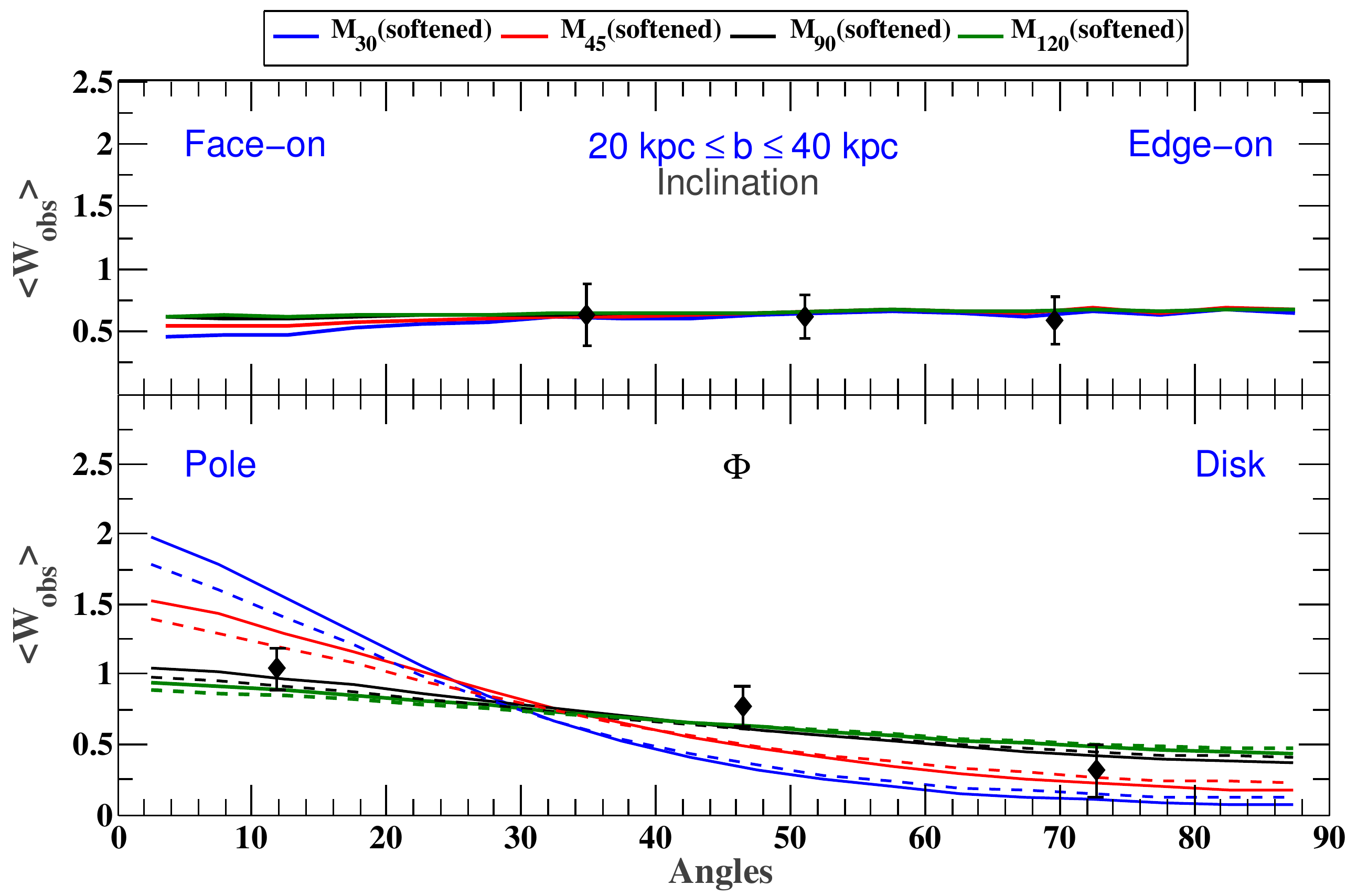}        
 \caption{The mean $\langle W_{obs} \rangle$ distribution for different models within $20 \leq b\leq 40$ kpc. The left panels show the sharp bipolar cone models, the sphere, the disk and the hybrid models that are combinations of the $M_{90}$ and the disk or the sphere. The right panels show the softened bipolar cone models.  The upper panels show the inclination profile, averaged over all azimuths. The lower panels are inclination-averaged azimuthal profiles. The (solid line) azimuthal profiles in both the left and right lower panels are averaged over i$\geq 45^{\circ}$. The dashed lines in the bottom right panel are averaged over all inclinations. The data points are adopted from the inclination and azimuthal profiles of disk galaxies in \cite{Bordoloi2011a}}.
\label{fig:model Wobs}
\end{figure*}

\subsection{$N_{HI}(i,\phi)$ column density distribution for different models}

The first observable we focus on is the distribution of absorption column density as a function of $i$ and $\phi$ at a fixed impact parameter. It can be used to predict the distribution of Mg II-selected galaxies in $(i,\phi)$ once a threshold is applied in equivalent width and after accounting for the $\sin i$ weighting in random orientations. 

For the purpose of presentation, we show these in $N_{HI}$ to avoid, for the time being, the question of saturation in Mg II.  Figure \ref{fig4} shows $N_{HI}(i,\phi)$ column densities for a set of sharp cones and the disk model at a fixed impact parameter, $b = 45$ kpc.  As expected, the spherical model M$_{180}$ has no variation of absorber strength with inclination and azimuthal angles. For all the other cases both the sharp bipolar cone models and the disk have the strongest absorption for edge on systems with $i$ approaching 90$^\circ$. However for bipolar cone models, strong systems are preferentially aligned near the minor axis of the galaxies (i.e. $\phi \rightarrow 0^{\circ}$) whereas for disk models this  is reversed and the strongest absorption is aligned along the disk of the galaxy (i.e. $\phi \rightarrow 90^{\circ}$).  It can be seen that all the models, except the completely spherical one, have regions where no absorption occurs. Quasar lines of sight with strong absorption line systems would not be observed in these regions of  $(i,\phi)$, while lines of sight without strong absorption (e.g. below 0.1 {\AA}) could be seen there.

\subsection{Azimuthal and inclination profiles}

By collapsing one of the angular variables, we can study how $N_{HI}$, and/or $W_{obs}$ depend on the other angular variable, averaged over some impact parameter range. These profiles can be compared with the observed profiles from the stacked spectra analysis of \cite{Bordoloi2011a}. To demonstrate how different model geometry is sensitive to $i$ and $\phi$, we consider all the 3-d and 2-d models as described below.

For each model $W_{obs}$ is computed as described in the previous section. We create two observables to compare with the stacked spectra (1) $\langle W_{obs} \rangle$ within $20 \leq b \leq 40$ kpc, averaged over all $\phi$, which gives the variation of $i$ within the $b$ bin. (2) $\langle W_{obs} \rangle$ within $20 \leq b \leq 40$ kpc averaged over all $i$, which gives the average variation of $\phi$ within the same $b$ bin. Figure \ref{fig:model Wobs} shows the inclination and azimuthal dependence for different sets of models. The left panel shows the fiducial models $M_{30}$, $M_{45}$, $M_{90}$, $M_{120}$, sphere and disk as described above.  The two hybrid models  Hyb1 ($M_{90}$ + sphere) and Hyb2 ($M_{90}$+ Disk) are also shown in this figure.  The inclination and radial profiles of these hybrid models are plotted on the left hand panel of figure \ref{fig:model Wobs}. The right panel of figure \ref{fig:model Wobs} shows the same profiles for softened bipolar cone models $M_{30}$(softened), $M_{45}$(softened), $M_{90}$(softened) and $M_{120}$(softened). 

The upper panels of figure \ref{fig:model Wobs} show the inclination profiles. As noted above, these show very little variation of absorption strength with inclination $i$. The bottom panels show the azimuthal profiles. On the left hand set of models, we show only those objects with $i \geq 45^{\circ}$ since this is where the azimuthal effects are strongest and we can directly compare these with the observations. On the right hand panel (with fewer models) we show both the azimuthal profiles for these inclined objects with $i \geq 45^{\circ}$(solid lines) and for all inclinations (dashed lines). The hybrid models (left panel) and the  softened model (right panel) both well represent the observations. The optimal opening angle and the optimal $\eta$ and $f$ for these set of models will be determined later in section 5. 

The bottom panel shows the azimuthal dependence of $\langle W_{obs} \rangle$ where a strong variation is seen. The solid lines show the profile for inclined galaxies ($i \geq 45^{\circ}$). The sharp bipolar cone models are always stronger along the pole of the galaxy ($\phi \sim 0^{\circ}$) whereas the disk model exhibits strongest absorption near the plane of the disk ($\phi \sim 90^{\circ}$).  Comparing with the data from \cite{Bordoloi2011a}, the disk model can be easily ruled out since the azimuthal dependence is of the opposite sign. It can be clearly seen from the top panel that the inclination dependence is rather a weak diagnostic of different model geometries and most of the models have essentially the same flat variation with inclination.  

It may strike the reader as surprising that there is such a weak dependence on inclination for the 2-d disk model.  However, there are two competing effects. The increased $\sec i$ path through the disk produces stronger absorption. On the other hand, as shown in equation \ref{eqn5},  lines of sight (except those that pass through the disk along the line of sights defined by the plane of the disk and the plane of the sky, i.e. those with $\phi = 90^{\circ}$ in equation \ref{eqn5} will intersect the disk at progressively larger distances as the inclination of the disk is increased, reducing the absorption. Which of these wins, i.e. the sign of the gradient of the overall $\langle W_{obs} \rangle (i)$ profile, depends on the radial profile of the disk, flatter profiles increase the absorption at high inclinations.

\begin{figure}
    \includegraphics[height=9cm,width=9cm]{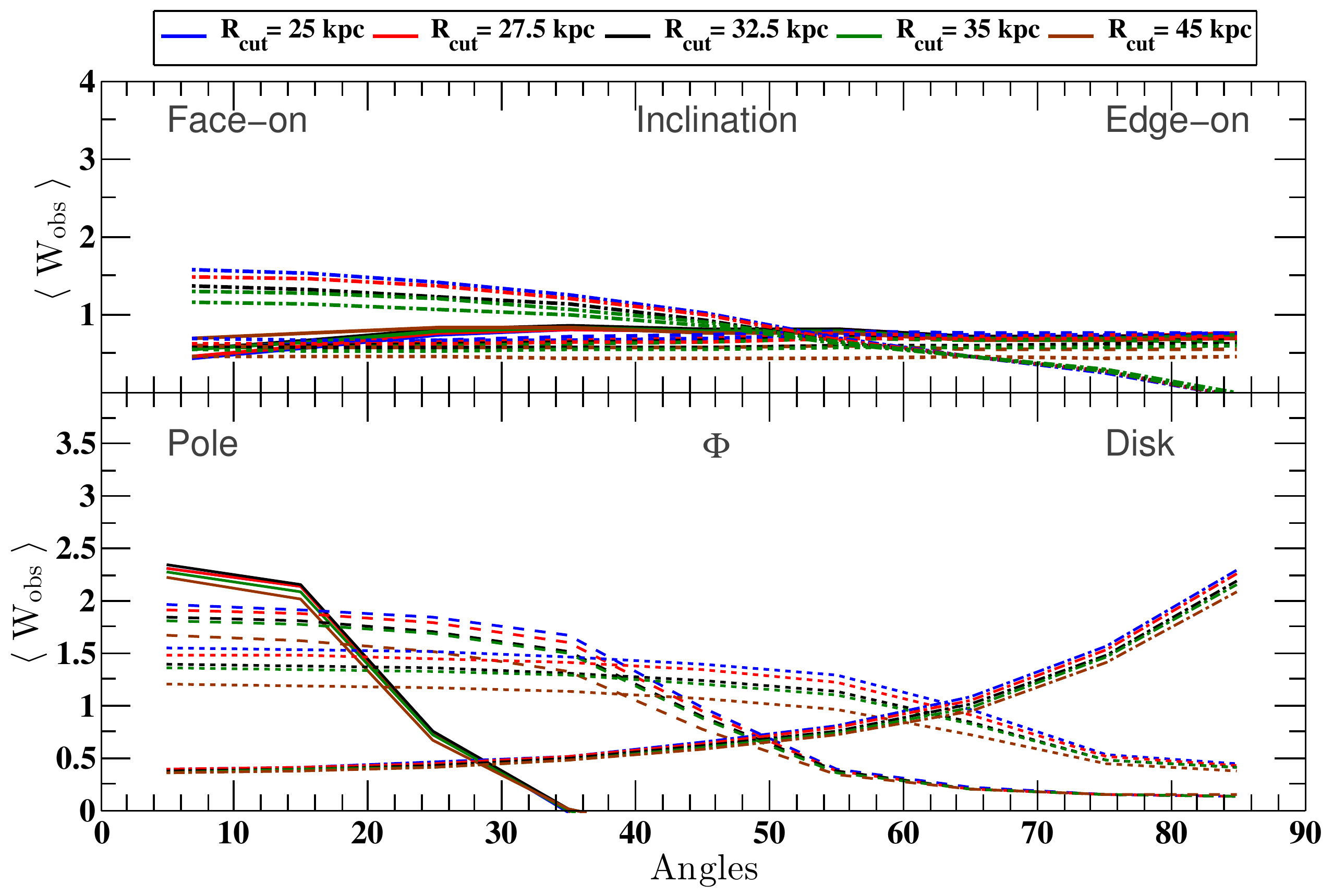}        
    \caption{The mean $\langle W_{obs} \rangle$ distribution for $M_{45}$ (solid lines), $M_{90}$ (dashed lines), $M_{120}$ (dotted lines) and pure Disk (dot-dashed lines) models which have different radial profiles obtained by setting $R_{cut}$ in the range $25 \leq R_{cut} \leq 45 $ kpc, computed as a function of inclination $i$ (top panel) and azimuthal angle $\phi$ (lower panel).  It is noticeable how the sign of the inclination dependence depends on the steepness of the radial profile, because of the two competing effects: the increased path length $\sec i$ vs. the effect of sampling greater distances in the disk - see text for discussion. The radial profile has little effect on the azimuthal dependence.}
\label{fig:model inclination difference}
\end{figure}

This dependence on radial profile is explored further in figure \ref{fig:model inclination difference}.  All the models in figure \ref{fig:model Wobs} have the same value of $R_{cut}$. We now investigate how the inclination and $\phi$ dependence vary when $R_{cut}$ is changed. Figure \ref{fig:model inclination difference} shows the variation in $\langle W_{obs} \rangle$ as a function of both $i$ and $\phi$ for the same baseline models if $R_{cut}$ is varied from 25 kpc to 45 kpc. Clearly the $\phi$ dependence remains much the same and each model can be clearly discriminated. The inclination dependence is much more sensitive to the selection of $R_{cut}$ rendering the discrimination between different models impossible. This plot highlights the fact that azimuthal dependence is a much stronger discriminant for Mg II absorber geometry.

\section{Application to Quasar Absorption Line Systems}

In this section we discuss how the model predictions described above can be translated into observables for a study done with quasar absorption line systems. We show how different models predict the covering fraction distribution for \emph{strong} ($W_{r}(2796)\geq 0.3$ {\AA}) absorbers, and study the distribution of azimuthal angles ($\phi$) of such absorbers in the data from \cite{Kacprzak2011b} and compare them to the expected distribution of different models.  


\begin{figure}
    \includegraphics[height=7cm,width=9cm]{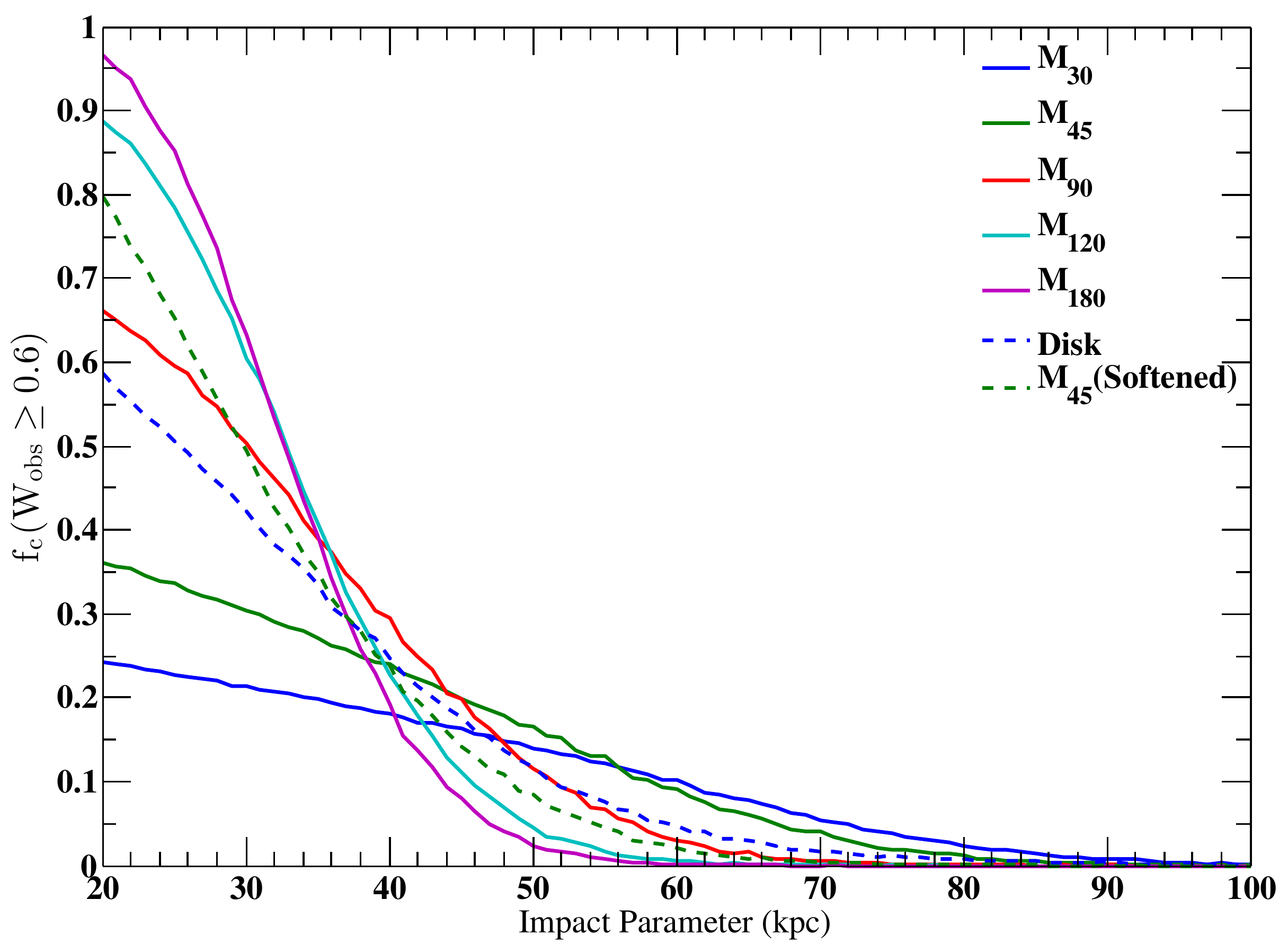}   
    \caption{The covering fraction for $W_{obs} \geq 0.6$ {\AA} absorbers as a function of impact parameter in different models, all based on idealized smooth 3-d density distributions.  The spherical model shows a sharp cutoff in covering fraction as expected. The behavior of the sharp bipolar cones vary with opening angle and approach the spherical case with increasing opening angle. The disk and the softened bipolar cone models exhibit different behavior but it is hard to distinguish between different models using this diagnostic.}
\label{fig: cov frac vs IP}
\end{figure}


\begin{figure}
    \includegraphics[height=7cm,width=9cm]{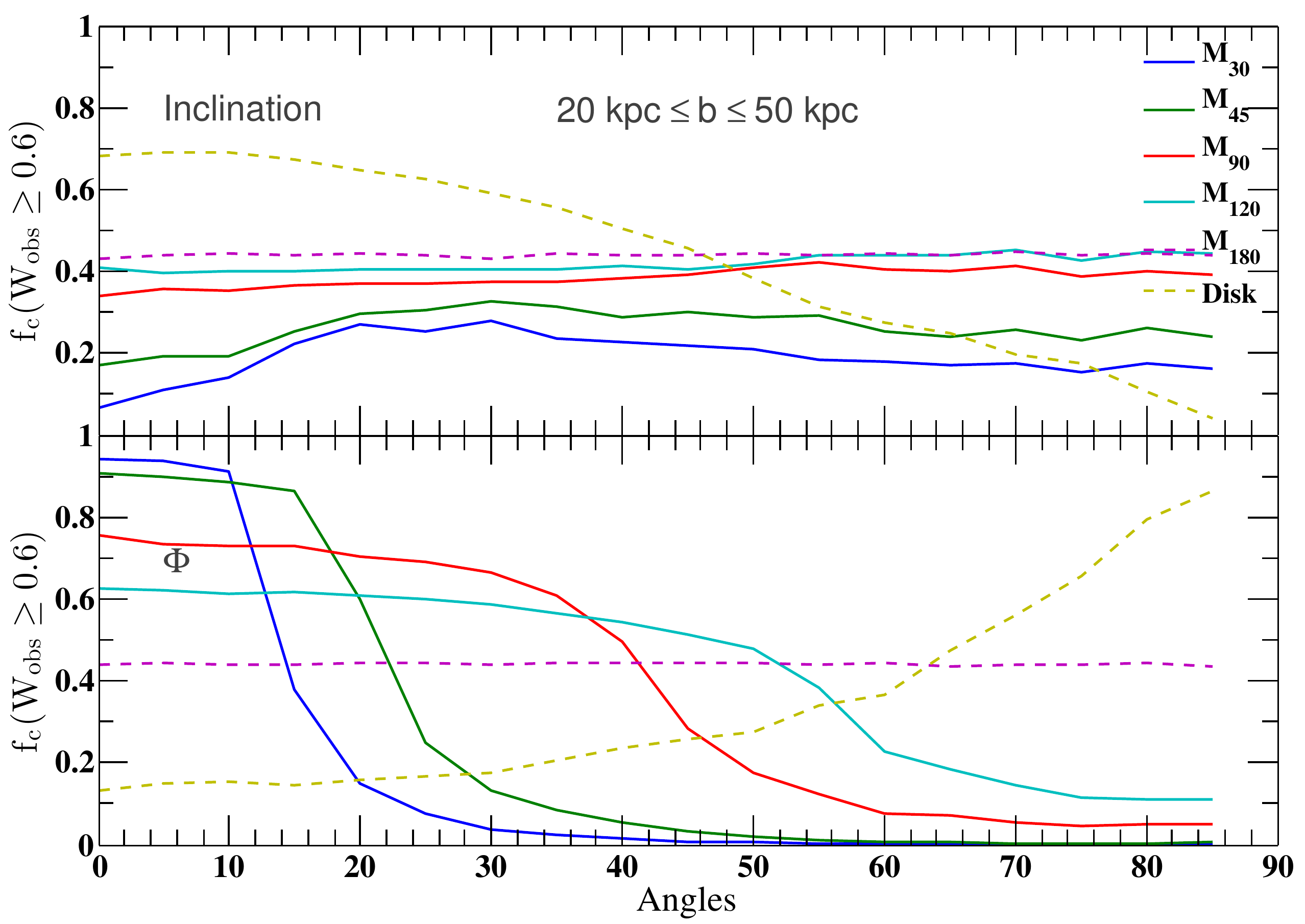}        
    \caption{Covering fraction for  $W_{obs} \geq 0.6$ {\AA} absorbers as a function of $i$ and $\phi$ for different models. Clearly the azimuthal dependence of covering fraction is a much stronger discriminant of absorber geometry although inclination dependence has a weak correlation with different models. }
\label{fig: cov frac vs angles}
\end{figure}

\subsection{Covering Fraction Distribution}
Extended Mg II absorbing gas is seen out to large impact parameters ($b > 100$ kpc) both from quasar absorption line studies and stacked spectra analysis. However both the absorption strength and gas covering fraction declines toward larger radii. Here, we first examine how the gas covering fraction $f_{c}(W_{obs} \geq 0.6$ {\AA}) varies with $b$ and $W_{obs}$.

In the context of our models, the gas covering fraction at an impact parameter $b$ is defined as 

\begin{equation}
f_{c}(W_{obs} \geq 0.6)(b) =  \frac{N_{( W_{obs} \geq 0.6)}(b)}{N_{tot}(b)}
\end{equation}

$N_{( W_{obs} \geq 0.6)}(b)$ is the number of lines of sight that have an absorber with equivalent width $W_{obs} \geq 0.6$ {\AA} at an impact parameter $b$ and $N_{tot}$ is the total number of lines of sight at the same impact parameter, given by the total number of orientations of the galaxy considered.  

Figure \ref{fig: cov frac vs IP} shows the covering fraction distribution of $W_{obs} \geq 0.6$ {\AA} absorbers as a function of $b$ for models $M_{30}$, $M_{45}$, $M_{90}$, $M_{120}$, Sphere, Disk and $M_{45}$(softened). This plot is similar to a covering fraction distribution of $W_{r}(2796) \geq 0.3$ {\AA} for a quasar absorption line study. But as the doublet ratio of  $W_{r}(2796) \sim 0.3$ {\AA} systems might vary between 1 and 2, there is indeed some leeway in this conversion and comparison. As expected the covering fraction for the Sphere model falls off sharply from almost unit covering fraction at low impact parameters to zero covering fraction as $b \rightarrow 50$ kpc. Although all the sharp bipolar cones have exactly the same average radial profile of total absorption, the covering fractions exhibit a range of radial profiles and approach the spherical case with increasing opening angle. While the disk and the softened bipolar cone models exhibit different behavior in detail, it is hard to distinguish between these different geometries purely based on their covering fraction as a function of impact parameter. For all the models, $f_{c}(W \geq 0.6) \rightarrow 0$ as $b > 80$ kpc. Observations indicate that the covering fraction does not go to zero as  $b > 80$ kpc, for quasar absorption line studies \citep{Nielsen2012}. But the models studied here are for isolated systems and at higher impact parameters contribution from filamentary accretion, satellite galaxies etc will become significant.

Clearly, the radial dependence of covering fraction is not very sensitive to model geometry. The variation of gas covering fractions with inclination and azimuthal angles offer greater discriminatory power. Figure \ref{fig: cov frac vs angles} shows the gas covering fraction for  $W_{obs} \geq 0.6$ {\AA} absorbers as a function of inclination angle ($i$) and azimuthal angle ($\phi$) within $20 \leq b \leq 50$ kpc. The $\phi$ dependence of covering fraction is quite obvious from this figure. The sharp bipolar cone models show very high covering fractions near the galaxy pole ($\phi \rightarrow 0$) and which gradually decrease as the cone opening angle increases. They approach the Sphere model as $\chi \sim 180^{\circ}$. The disk model shows the opposite trend and has a very high covering fraction along the disk axis. The inclination dependence is not as strong as the azimuthal dependence. Only the 2-d disk model exhibits higher covering fraction for the face-on galaxies ($i=0^{\circ}$) as compared to the edge-on galaxies ($i=90^{\circ}$). 

It should be stressed that any comparison of these covering fraction estimates with the observations should be done with care. These covering fraction estimates show how different Mg II absorption distribution geometries manifest themselves in terms of the observed covering fractions relative to each other. The models studied here represent isolated galaxies and the CGM is a more complex environment. These models of isolated systems do not include the effects of filamentary accretion, satellite galaxies or LMC like streams or high velocity clouds (HVCs). Further, we assume a smooth distribution Mg II absorption, and in reality the CGM is more patchy. At high impact parameters these structures will start to contribute significantly to the overall covering fraction. Hence the comparison with the observed covering fraction should be done with these caveats in mind. 


\begin{figure*}
   \centering
    \includegraphics[height=9cm,width=11.5cm]{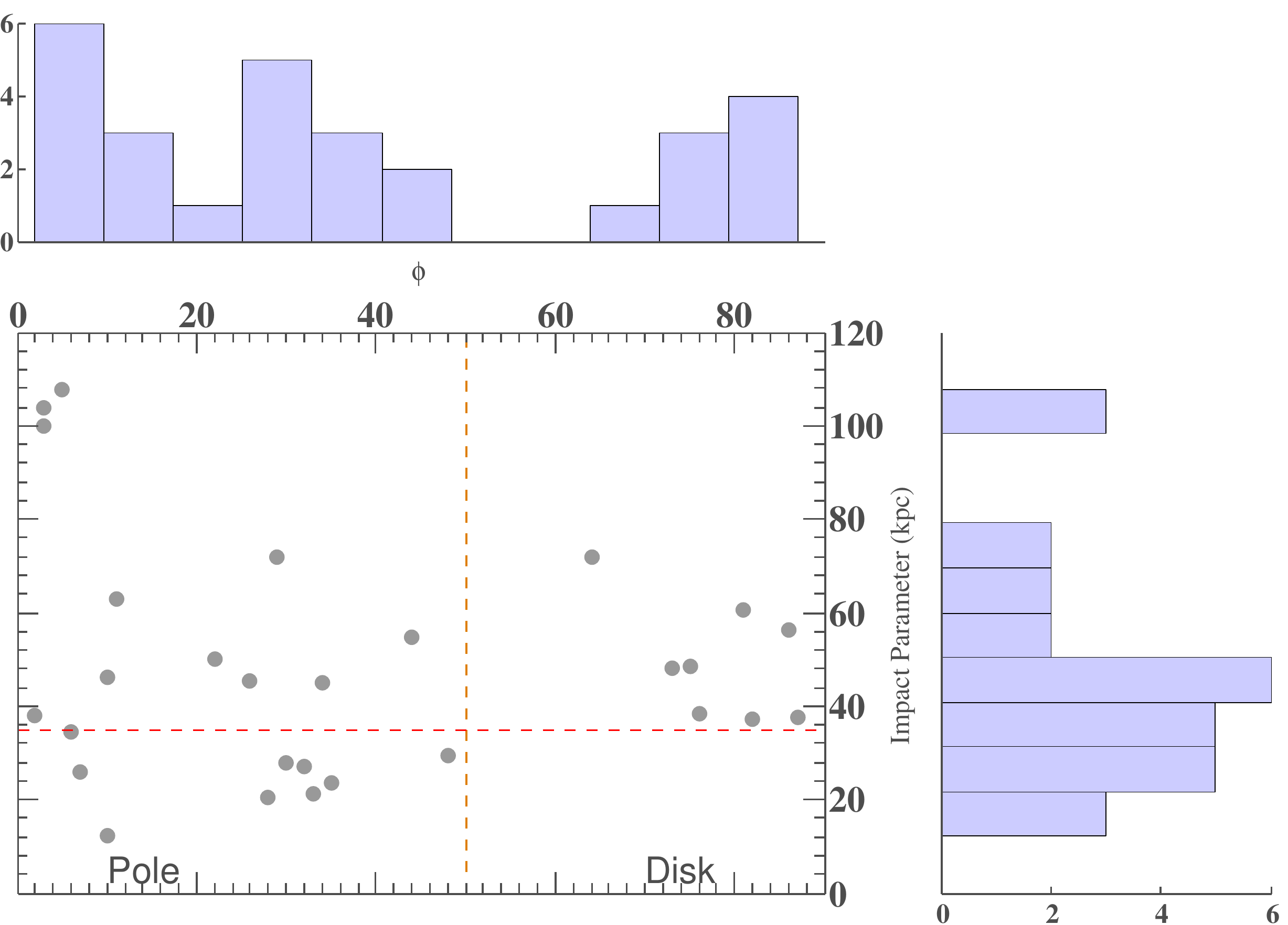}        
    \caption{The distribution of ``strong'' Mg II absorbers ( $W_{r}(2796) \geq 0.3 $ \AA) in azimuthal angle and impact parameter $b$ (adopted  from \citealt{Kacprzak2011b}). 72\% of the absorbers lie within $50^{\circ}$ of the semi minor axis of the galaxies, i.e. lie to the left of the vertical dashed line. The horizontal dashed line marks $b \leq$ 35 kpc. Within this distance, \emph{all} of the absorbers lie within $50^{\circ}$ of the semi-minor axis of the galaxy. }
\label{fig: quasar angles}
\end{figure*}



\begin{figure}
    \includegraphics[height=7cm,width=8cm]{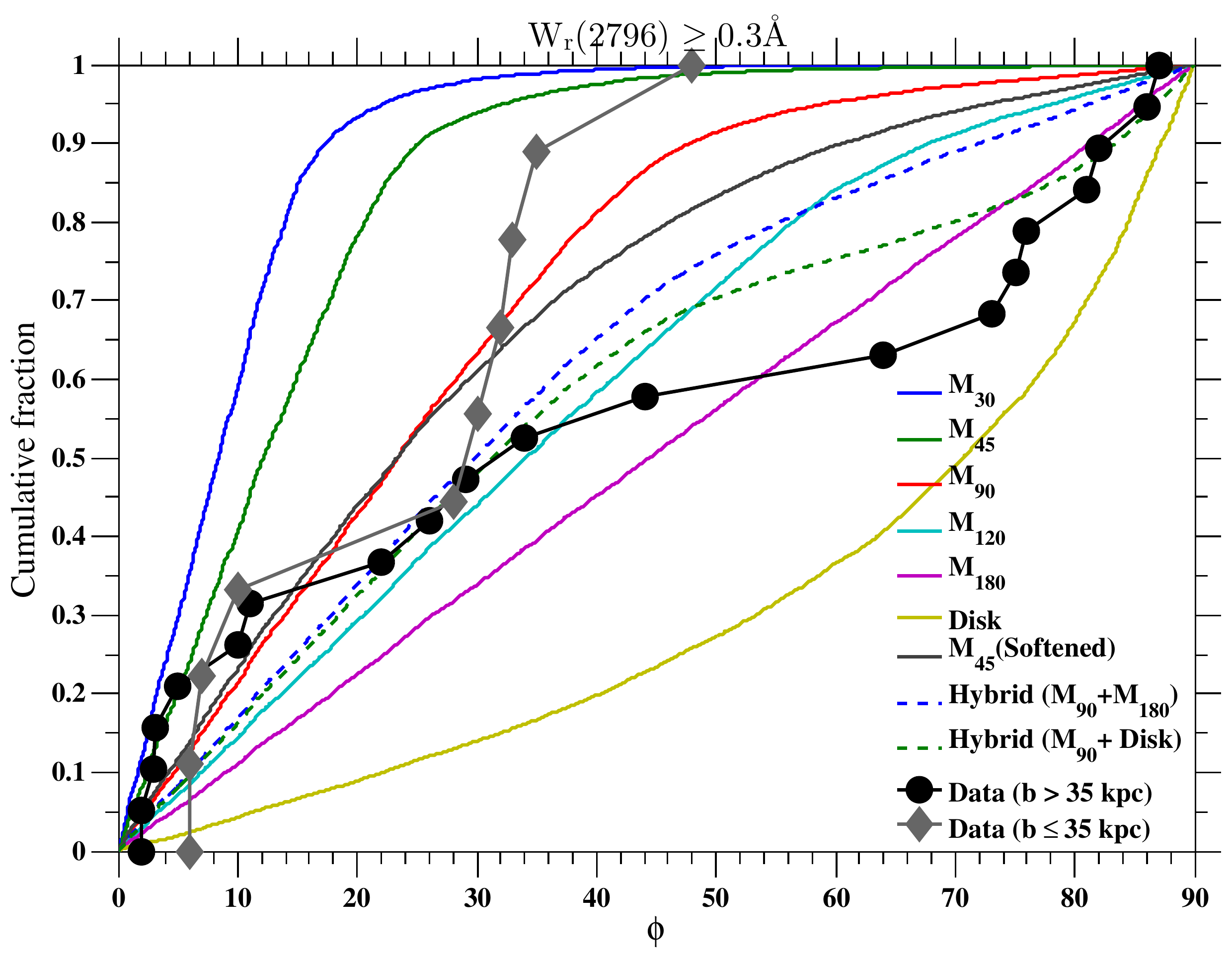}        
    \caption{Cumulative distribution of azimuthal angles ($\phi$) moving away from the projected minor axis for different 3-d models and for the quasar absorption lines with ($W_r(2796)\geq$ 0.3 {\AA}, adopted from \citep{Kacprzak2011b}. The data clearly rule out a spherical or a disk-like geometry alone. However a hybrid cone+disk or cone+sphere model or a softened bipolar cone structure is consistent with the data, depending on the opening angle of the cone.}
\label{fig: cdfplot}
\end{figure}

\subsection{Distribution of Azimuthal Angles in quasar data}

In this section we discuss the azimuthal dependence of a set of strong ($W_r(2796) \geq 0.3$ {\AA}) Mg II absorbers observed in \cite{Kacprzak2011b}. \cite{Kacprzak2011b} had a sample of 40 Mg II absorbing galaxies, out of which 29 have a rest frame equivalent width $W_r (2796) \geq 0.3$ {\AA}. One out of these 29 galaxies is completely face on ($i \leq 5^{\circ}$) and it becomes rather difficult to assign an accurate azimuthal angle. Hence we exclude this galaxy from our analysis. We shall focus our analysis on this sample of 28 galaxies classified as ``strong'' absorbers. Galaxy morphology, inclination and azimuthal angles were obtained by using two-dimensional fits of a (co-centered) bulge and disk model. We refer the reader to \cite{Kacprzak2011b} for details of methodology and data reduction details.

Figure \ref{fig: quasar angles} shows the distribution of ``strong'' Mg II absorbers as a function of their azimuthal angle ($\phi$) and impact parameter (b). 72\% of all the ``strong'' absorbers (20 out of 28) that are within $b\; < 120$ kpc lie within $50^{\circ}$ of the semi minor axis of the galaxies i.e. to the left of the dashed vertical line of figure \ref{fig: quasar angles}, as compared with 55\% that would be expected for a uniform azimuthal distribution.  More significantly, for $b\; \leq$ 35 kpc (below the horizontal dashed line) all of the nine ``strong'' absorbers lie within  $50^{\circ}$ of the galaxy semi minor axis. This evidence for a strong azimuthal dependence supports the hypothesis that out to $b\; \sim$ 40 kpc the ``strong'' Mg II absorbers are originating in bipolar galactic winds, as indicated by the measurement of total Mg II absorption in \cite{Bordoloi2011a}. The lack of absorber-galaxy pairs between $\phi \; \sim$ 50 to 65 degrees is statistically significant. For a random distribution of $\phi$  the probability of having no absorber-galaxy pairs within $\sim$ 20 to 65 degrees is 0.6$\%$, computed using $10^{6}$ simulated samples (with $N_{pair} = 28$), drawn from a uniform random distribution. Hence the lack of absorber-galaxy pairs seen within 20-65 degrees in figure \ref{fig: quasar angles} is significant at $\geq$ 2.5-$\sigma$ level.

To compare this dataset with models we plot the cumulative distribution of $\phi$ of the Mg II-selected galaxies and a set of models in figure \ref{fig: cdfplot}.  The gray squares represent the cumulative distribution function for the absorbers detected at $b \leq 35$ kpc. Clearly this data favor a more cone like distribution of $\phi$. A two sample KS test rules out the null hypothesis that the distribution of azimuthal angles of the disk model, $M_{30}, M_{45}$ and the sphere model and the low b data are drawn from the same parent distribution at 1\% significance level.  Even for the absorbers at $b > 35$ kpc (black circles), the quasar absorption line data immediately rule out the disk model at 1\% significance level.  Also $M_{30}$ and $M_{45}$ models are ruled out by the data at 1\% significance level by a two sample KS test. The data can however be reproduced by the softened bipolar cone model or a hybrid model.  At low impact parameters, the data are consistent with models of small opening angle whereas the high impact parameter data are consistent with high opening angle models. In the next section, we quantify the models supported by the quasar absorption data as well as the integrated spectra data with a 2-D KS test.

\section{Joint Analysis of Stacked Spectra and Quasar Absorption Line Systems}
\label{sec:joint_analysis}
In this section, we carry out a joint analysis of the two independent and \emph{orthogonal} probes of Mg II absorbers discussed above, namely the stacked spectra data of \cite{Bordoloi2011a} and the quasar absorption line probes from \cite{Kacprzak2011b}.

We chose a grid of models with opening angles within $1^{\circ} \leq \chi \leq 170^{\circ}$ for both the sharp bipolar cone and the softened bipolar cone models. We create, as above, two sets of hybrid models combining a sharp bipolar cone model first with a sphere model and second, with a disk model as described in the previous Section 2.4. We examine the  cone+sphere hybrid model in the over-density (equation \ref{density_ratio}) range of $0.1 < \eta < 10$ and the cone+disk model in the range of $0.1 \leq f \leq 10$ (equation \ref{whyb}). For each model the overall normalization factor $\alpha$ is assigned so that  $\langle W_{obs} \rangle \sim 0.7 $ {\AA} within $20 \; \rm{kpc} \leq b \leq 50$ kpc.

To compare the models with the integrated spectra data, we compute the mean azimuthal profile for each model within a given bin of impact parameter. The mean azimuthal profiles of the models and the measurements are compared by computing the $\chi^2$ value for each model, taking the minimum $\chi^2$ value as the best fit parameters for each particular set of models. 

The quasar absorption line data are compared with the same set of models in the same $b$ bins by performing a two dimensional Kolmogorov-Smirnov test \citep{Peacock1983,Fasano1987,Press_numerical_recipe}. To perform the two dimensional Kolmogorov-Smirnov test we chose all ``strong'' absorbers (above 0.3 \AA) within a fixed $b$ range and compare the distribution of inclination and $\phi$ angles with that expected for different models. 

\begin{figure}
\centering
    \includegraphics[height=7cm,width=8cm]{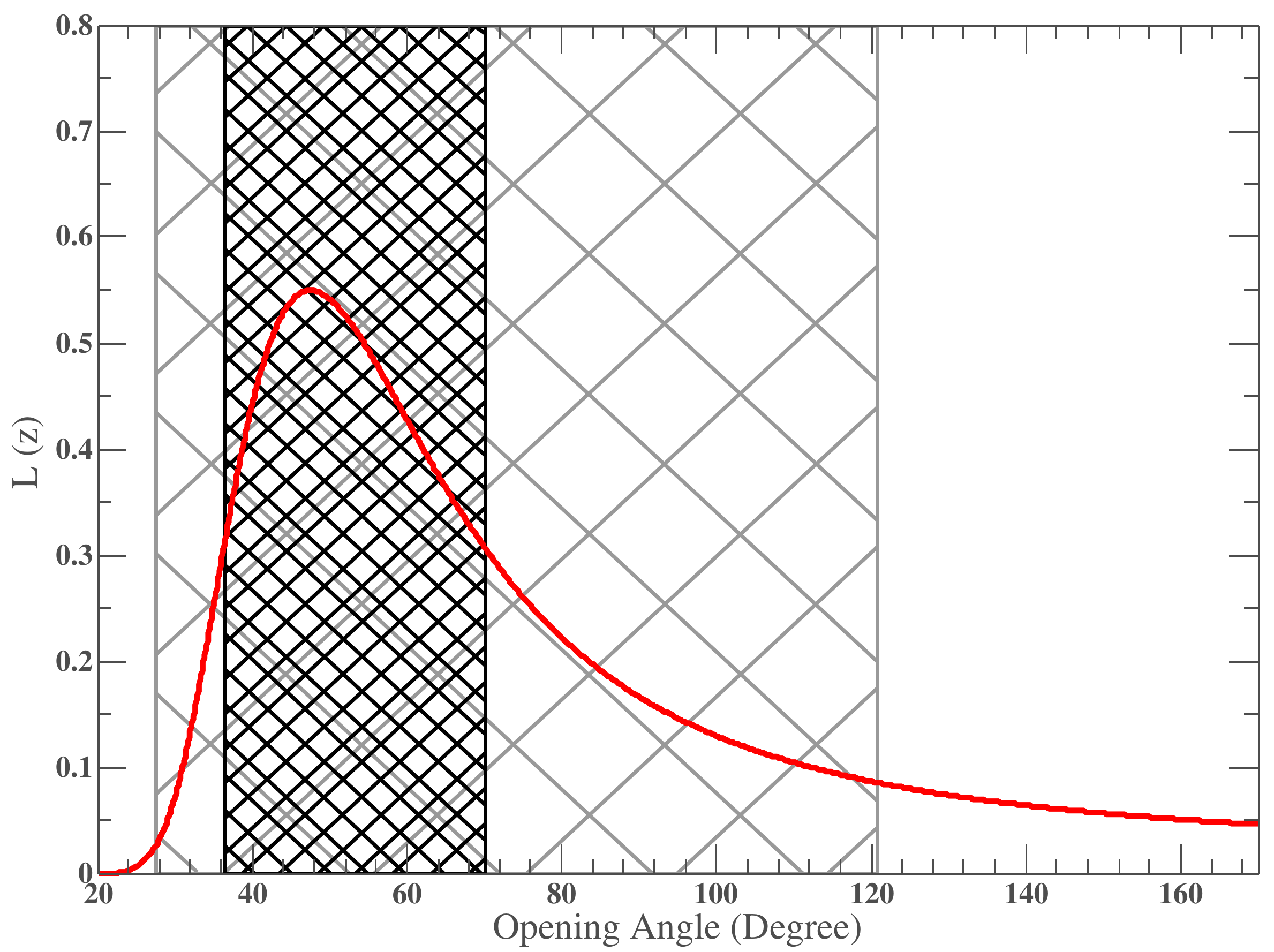}        
    \caption{The joint constraints on the opening angle of the softened bipolar cone model at $20 < b < 40$ kpc from both the stacked spectra analysis and the quasar absorption line studies. The red solid line is the normalized likelihood curve obtained from the $\chi^2$ minimization of the model predictions to the azimuthal profile in the stacked spectra analysis. The maximum likelihood opening angle of about $45^{\circ}$ is consistent with the opening angles allowed by performing a 2D KS test of the model angle distributions with the quasar absorption line data (shown as the open hashed region).  The close hatched region shows the joint constraint at 90\% confidence level.}
\label{fig: likelihood}
\end{figure}

To estimate the model parameters which best represent the observations, we compare the observations with the model predictions within $b \leq 40$ kpc. We first focus on the softened bipolar cone model and compare the models within $20 \; \rm{kpc} \leq b \leq 40\; \rm{kpc}$ with the observations. Figure \ref{fig: likelihood} shows the joint constraints from both the integrated spectra data and the quasar absorption line data. The red curve shows the likelihood curve from the integrated spectra data and the gray hashed region gives the opening angles allowed by the 2-D KS test at 10\% significance level (i.e. parameters outside of this region are excluded with more than 90\% confidence). The black hashed region is the joint constraints from both these probes which are obtained by simply taking the product of the probabilities. This analysis gives an opening angle of 45$_{-9}^{+15}$ degrees that is allowed by both the observations for a softened bipolar outflow model at 90\% confidence.

\begin{figure*}
\centering
 \includegraphics[height=7.5cm,width=8cm]{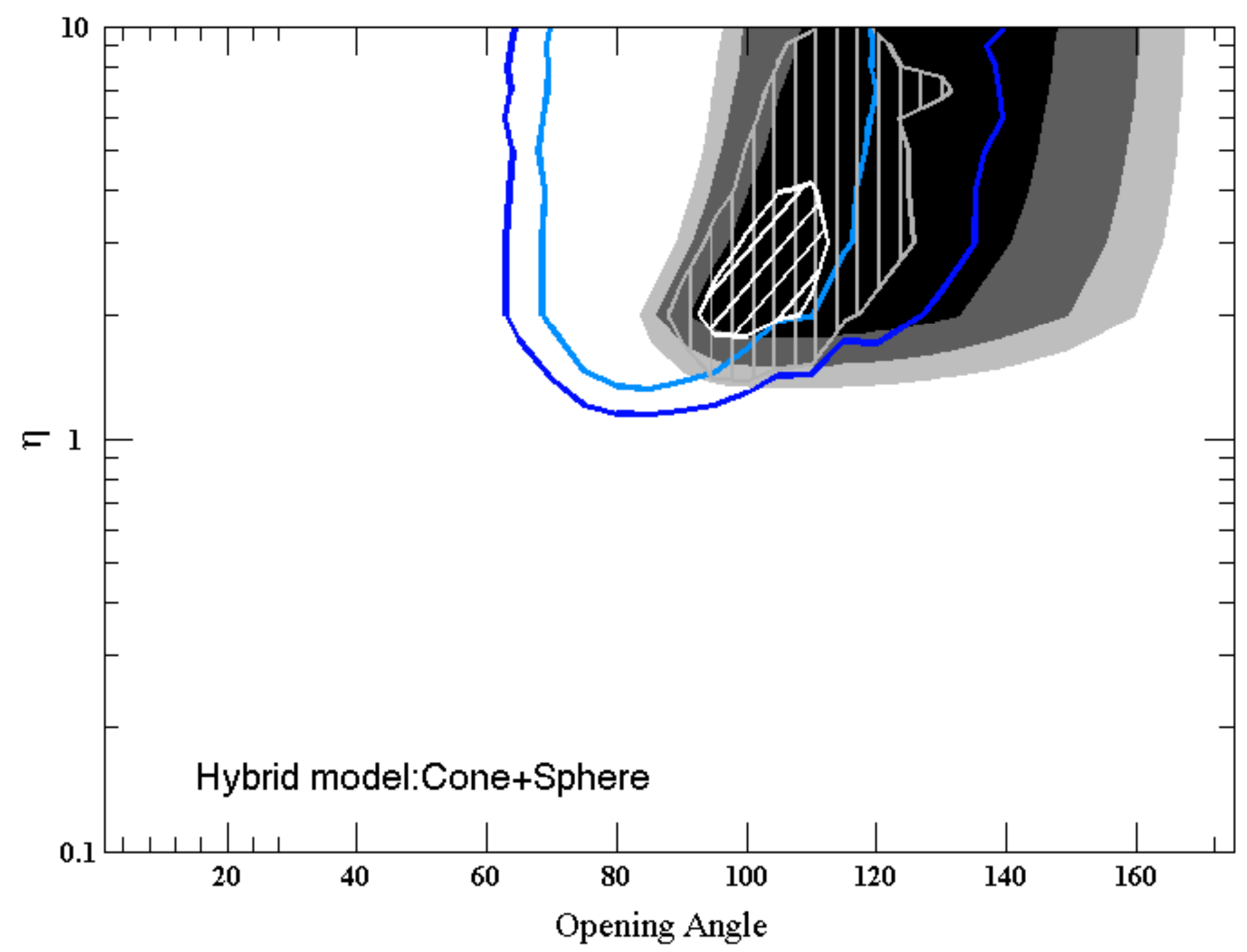} 
 \includegraphics[height=7.5cm,width=8cm]{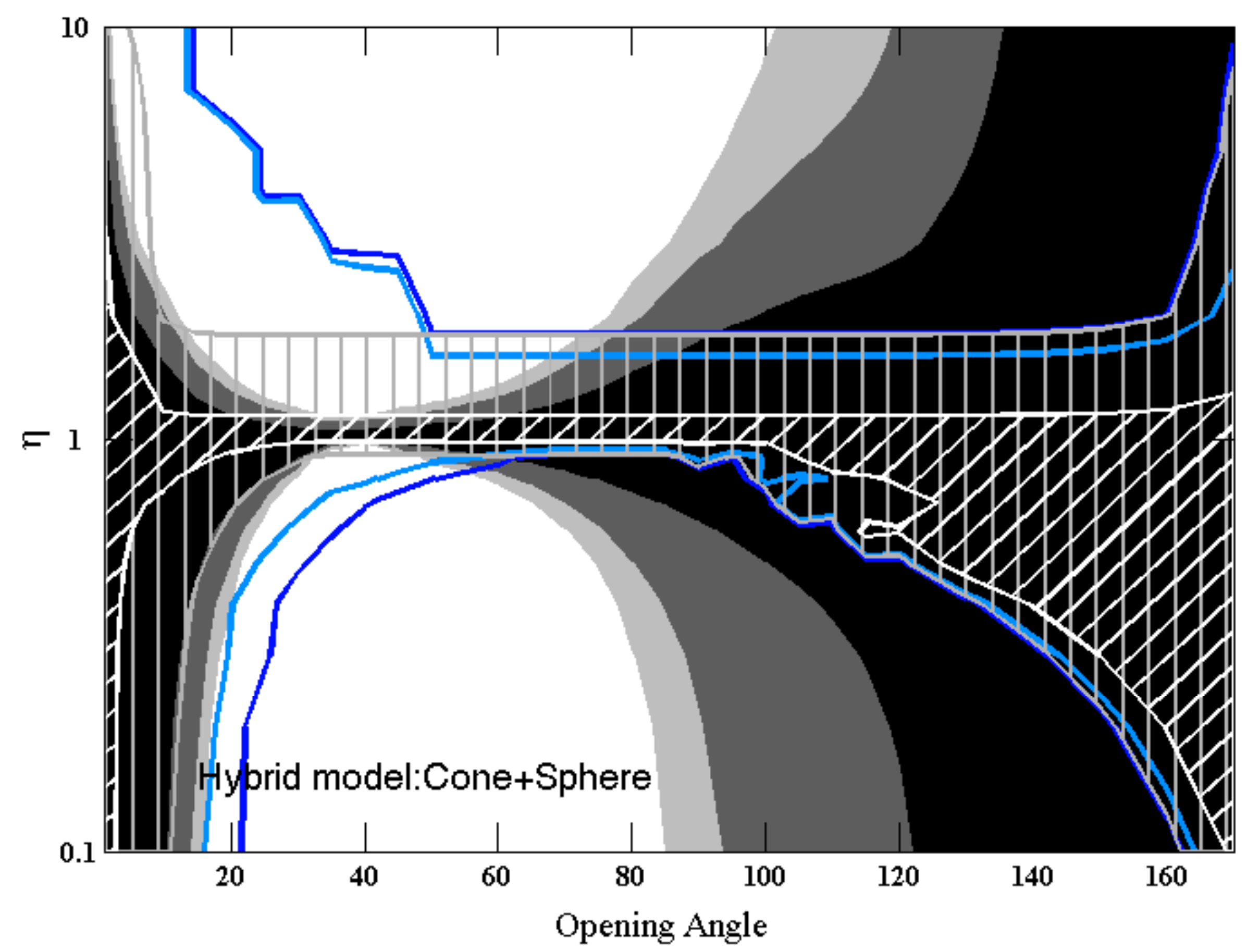}
 \caption{Joint constraints on the hybrid Cone+Sphere model from both stacked spectra analysis and the quasar absorption line studies on two parameters describing this model - these are $\eta$, the density contrast above the pole and in the plane of the disk, and the cone opening angle. The 68\%, 90\% and 95\% confidence contours obtained from $\chi^2$  fitting of the model predictions to the azimuthal profile in the stacked spectra analysis (gray contours). The blue contours show the parameter space allowed by the quasar absorption line data at 90\% and 95\% confidence level, obtained from a 2D KS test analysis. The hashed region on both panels show the parameter space allowed by both datasets at 90\% and 95\% confidence level. The left panel is for absorption within $20\;\rm{kpc} \leq b \leq 40$ kpc, and the right panel is for absorption within $40\;\rm{kpc} \leq b \leq 80$ kpc. The differences show the pronounced weakening of the azimuthal asymmetries with radius.}
\label{fig: Hybrid_confidence}
\end{figure*}

We now shift our focus on the two hybrid models. We first examine the hybrid model having a cone and a sphere component. As described above we compute the $\chi^2$ values, by comparing each model with the integrated spectra. The allowed confidence contours for the two parameters of interest, $n$ and opening angle are shown in figure \ref{fig: Hybrid_confidence}. The contours show the 68\%, 90\% and 95\% confidence intervals of the allowed model parameters. The blue contours are allowed regions in parameter space obtained from the quasar absorption line data at 95\% and 90\% confidence level. The hashed regions give the joint constraints on opening angle and $\eta$ at 95\% and 90\% confidence levels respectively. The left panel shows the allowed model parameters within $20 \; \rm{kpc} \leq b \leq 40\; \rm{kpc}$. At these impact parameters, both the integrated spectra data and the quasar absorption line data favor models where there is an enhancement of Mg II absorption within $\sim$ 100$^{\circ}$ of the bipolar region. The opening angle allowed by both the datasets within 40 kpc is 105$_{-10}^{+5\;\circ}$ and the over-density $\eta$ is 2$_{-0.2}^{+1.3}$. 

\begin{figure*}
\centering
 \includegraphics[height=7.5cm,width=8cm]{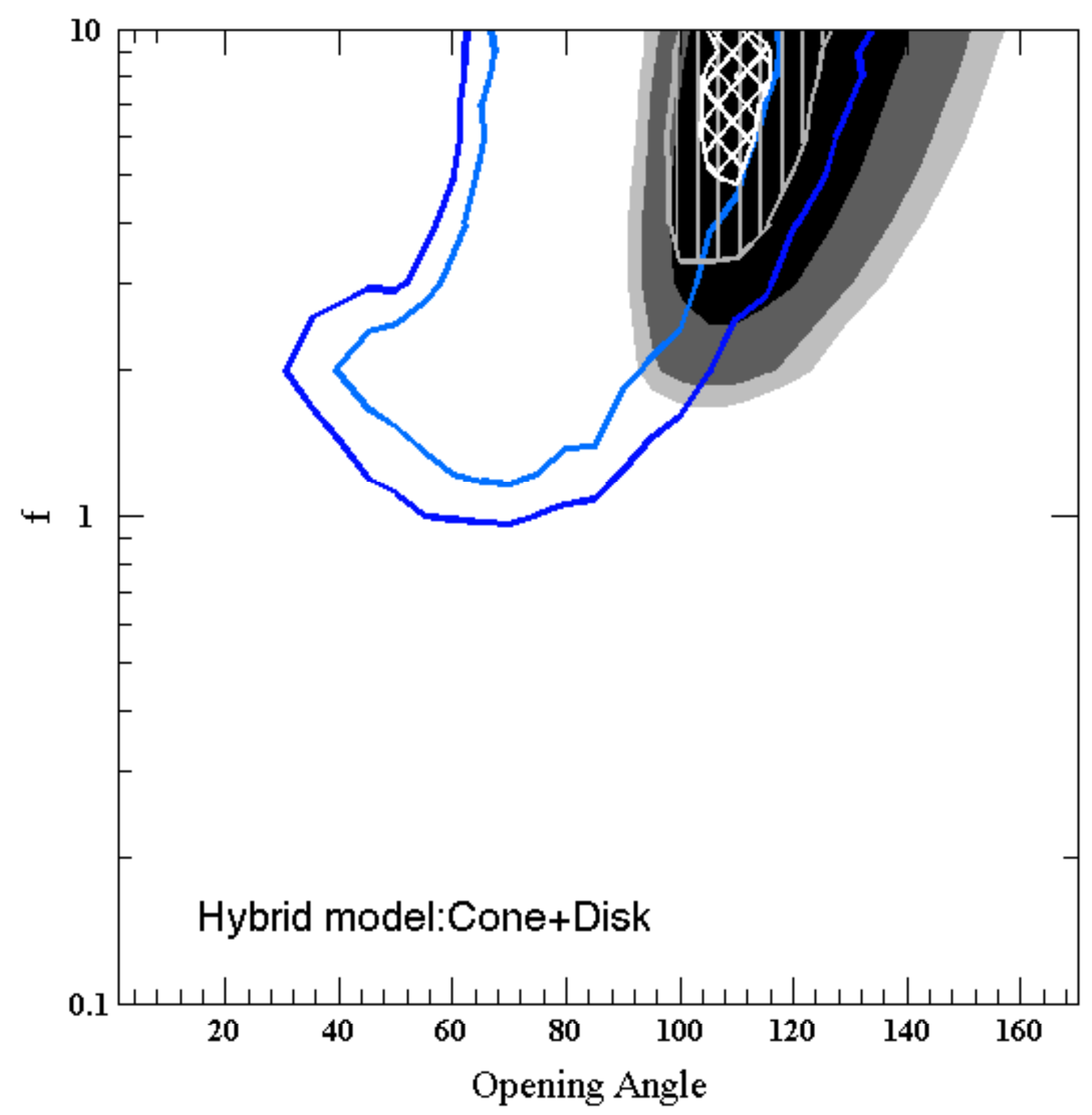}
 \includegraphics[height=7.5cm,width=8cm]{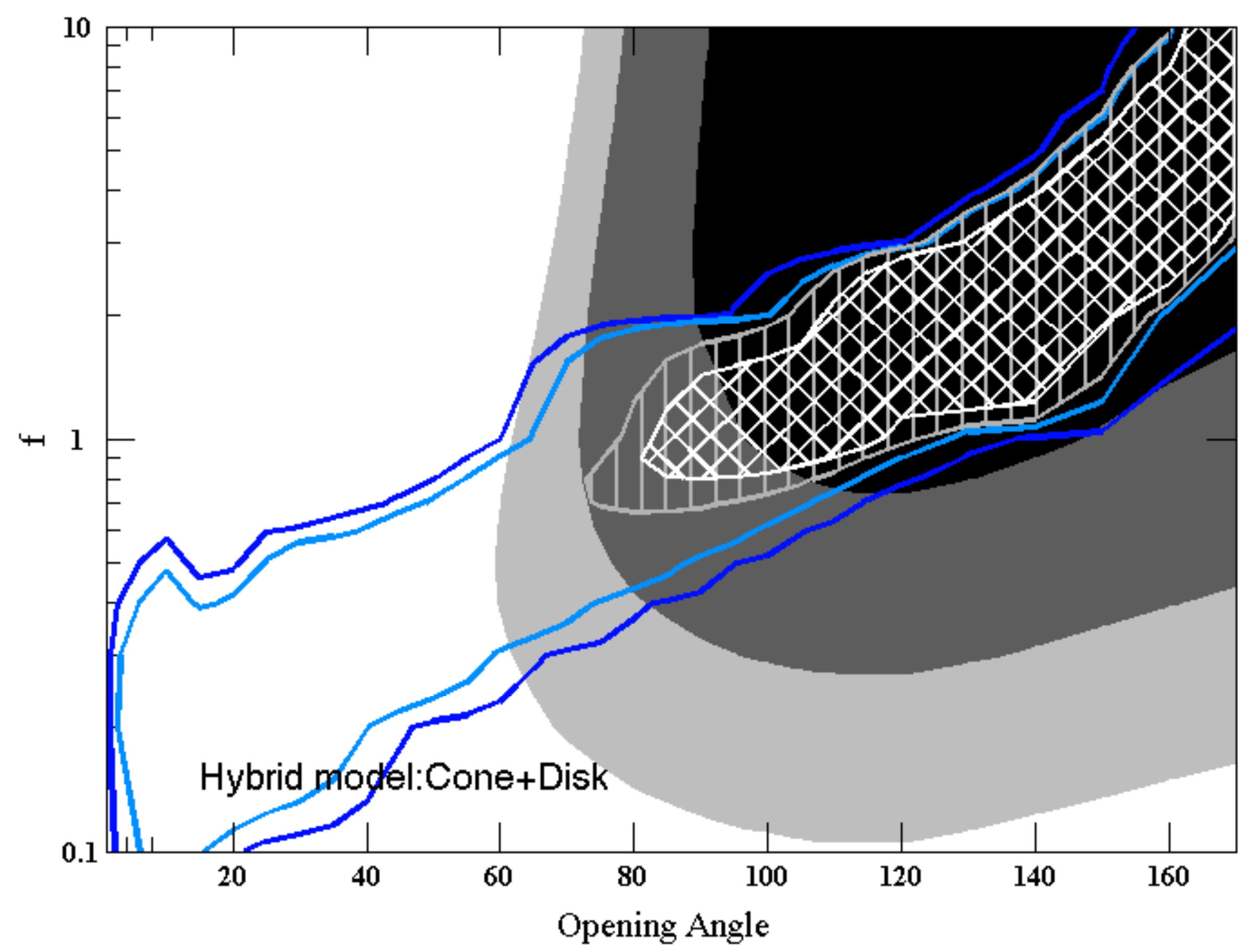}        
 \caption{As in Figure \ref{fig: Hybrid_confidence}, but for the hybrid Cone+Disk model.  The parameters are now $f$, the ratio of the total amount of material in the cone to the disk, and the cone opening angle.  }
\label{fig: Hybrid_confidence2}
\end{figure*}

We now look at the second hybrid model having a cone and a disk component. Figure \ref{fig: Hybrid_confidence2} shows the constraints on the allowed model parameters, the plotting scheme is the same as described above. The close in data (left panel, $20 \; \rm{kpc} \leq b \leq 40\; \rm{kpc}$) show that both the quasar absorption line systems and the integrated spectra are in agreement with a cone like geometry with an opening angle of $\sim$ 110$^{\circ}$. The quasar absorption line data allow for a range of opening angles which are smaller than those allowed by the integrated spectra data. We put joint constraint on the opening angle at 110$^{\circ}\pm 5 ^{\circ}$ and the fractional contribution to the cone component is found to be $f= 6^{+4}_{-1}$.

The Mg II absorption around galaxies show strong azimuthal asymmetry for $b < 40$ kpc. But at higher impact parameters, this azimuthal asymmetry is much weaker as shown in \cite{Bordoloi2011a}. We now investigate the models that can best represent the observations within $41 \; \rm{kpc} \leq b \leq 80\; \rm{kpc}$ as shown in the right panel of figure \ref{fig: Hybrid_confidence}. We see that the models with $\eta=1$ i.e. spherical distribution of Mg II gas around the galaxies, cannot be ruled out by either of the datasets.  Both the datasets are insensitive to the very low and high opening angles since such very sharp features produce little effect on the overall model. By combining the two probes together we see that the spherically symmetric distribution with $\eta \sim 1$ is strongly preferred. 

Furthermore, the joint models obtained for the two impact parameter ranges at 90\% confidence do not overlap.  This is evidence that the distribution of Mg II absorbers changes from a highly asymmetric distribution at low impact parameters to a less asymmetric distribution, consistent with a spherical distribution, at high impact parameters.

A similar conclusion is reached from considering the second hybrid model (cone+disk) at the larger impact parameters (right panel, figure \ref{fig: Hybrid_confidence2}).  The quasar absorption line data are quite degenerate in $f$ and opening angle (requiring $f$ to reduce as the opening angle is narrowed) but are consistent with the stacked galaxy analysis. For allowed small opening angles there is significant contribution from the disk component. As the allowed opening angle increases the cone component becomes more and more dominant. Combining these two probes, we get joint constraints on the opening angle = 130$_{-45}^{+40\; \circ}$ and  $f = 2^{+8}_{-1.1}$. This best fit model also represents an almost spherical geometry with very large opening angle for the cone component. As before, the region allowed at small impact parameters does not overlap with the region allowed at larger radii.  

We conclude that the distribution of Mg II gas at low impact parameters is not the same as that found at high impact parameters. This radial trend is hard to see on either the integrated spectra or the quasar absorption line data alone because individually, either probe is not sensitive enough to rule out the same regions of parameter space at different $b$ (figure \ref{fig: Hybrid_confidence}, \ref{fig: Hybrid_confidence2}).  However the joint analysis reveals that the Mg II gas is distributed preferentially along the minor axis of disk galaxies with an opening angle of $\sim$ 100$^{\circ}$ and an overdensity of $\eta \sim 2-3$  and $f \sim 6-7$ within $b \leq 40$ kpc. At high impact parameters the distribution is no longer aligned only along the minor axis and a spherical distribution is also consistent with both the dataset.

These models quantify the radial changes in the distribution of Mg II absorbers around galaxies. With the best fit models we can present a more clearer picture of the radial distribution of Mg II absorbing gas. At close impact parameters (e.g. $b \leq 40$ kpc), the Mg II absorption is primarily aligned along the minor axis of the galaxy which suggest that they might be arising due cool to material entrained in star formation driven outflows. As we move outwards, the distribution of the absorbers are consistent with either a spherical distribution or a cone with a large opening angle augmented by an extended disk component which might be the signature of in falling material, perhaps as part of a galactic fountain or a contribution from material falling in from larger distances. The more spherical distribution could also reflect the contribution of satellite galaxies.




\section{Conclusion and Discussions}

In this paper we address the principle question of the distribution of cool Mg II gas observed as absorption line systems in the spectra of background sources. We present a suite of 3-dimensional models of Mg II absorption to study the azimuthal and inclination dependence of observational quantities, including the integrated column density and the probability of intersecting a region above a given column density threshold.  We compare these model distributions to observations of Mg II absorption that we have previously obtained by stacking the spectra of background galaxies \citep{Bordoloi2011a} and by analyzing, in this paper, the distribution Mg II absorbers found in quasar spectra as a function of the impact parameter and relative galaxy orientations \citep{Kacprzak2011b}. Both these \emph{orthogonal} probes of cool ionized Mg II gas, allow us to put joint constraints on the distribution of MgII gas. The main findings of this analysis can be summarized as follows.

 \begin{enumerate}
 
 \item The azimuthal dependence of the observable quantities is a much stronger diagnostic of the 3-d geometry than the dependence on inclination. 
 
 \item  A composite model that consists of a simple bipolar cone plus a spherical or disk component or a single softened bipolar distribution can well represent the azimuthal dependencies observed in \cite{Bordoloi2011a} at $b < 40$ kpc.  
  
 \item The strong Mg II absorbers observed in \cite{Kacprzak2011b} are also asymmetrically distributed. Overall, 72\% of these the absorbers are within $50^{\circ}$ of the galaxy semi minor axis. Within $b \leq 35$ kpc, ``all'' the strong absorbers are within  $50^{\circ}$ of the galaxy semi minor axis.  Comparing these absorbers with models and plotting the cumulative distribution of the azimuthal angles, we see that the data is consistent with the distribution of azimuthal angles of bi-polar models and any purely spherical or disk model can ruled out at low impact parameters.
 
 \item A joint analysis combining the integrated spectra data and the quasar absorption line data shows that within 40 kpc both these datasets can be well reproduced by a softened bipolar cone model of opening angle 45$_{-9}^{+15}$ degree or hybrid models consisting of a cone and a sphere  component with $\eta$ is 2$_{-0.2}^{+1.3}$ and an opening angle of 105$_{-10}^{+5}$ degree. A second hybrid model with a cone and a disk component is also consistent with the observed data with an opening angle of 110$^{\circ}\pm 5 ^{\circ}$ and $f= 6^{+4}_{-1}$. 
 
 \item These analyses show that the strong Mg II absorbers at low impact parameters are primarily distributed along the minor axis of disk galaxies and suggesting that they are perhaps primarily originating in galactic winds, with only a small fraction might be aligned along the major disk axis.  In rough terms, the number density of MgII above the poles of the galaxies are 2-3 times higher than in the planes of the disks.
  
 \item At larger impact parameters ($41\;\rm{kpc} < b < 80$ kpc), the distribution of Mg II gas is more symmetric than that observed at smaller impact parameters. A joint analysis of the integrated spectra and the quasar absorption line data shows that the model best representing both the datasets is a more or less spherical distribution of Mg II gas. 
 
 \item The clear change in the distribution of Mg II absorption with increasing impact parameter suggests that the origin of the absorption may also change as we move out in radius from the galaxies. The more symmetric distribution at large radii may reflect an increased contribution from in-falling material, or even satellite galaxies.
 
 \end{enumerate}
 
We also present predictions for the expected covering fraction of strong absorbers as a function of impact parameters for different absorber geometries. The covering fraction as a function of impact parameter is not a strong discriminant of absorber geometry.

We predict the expected inclination and azimuthal dependencies for individual QSO sight lines within $20$ kpc $\leq b \leq 50$ kpc for a set of given models. The covering fraction of Mg II absorbers are most sensitive to galaxy azimuthal angles. Amongst the covering fraction profiles, the azimuthal profile of gas covering fraction is the most sensitive probe to discriminate different absorber geometries.

This work has been supported by the Swiss National Science Foundation.  We also acknowledge useful discussions with Nicolas Bouch\'{e}.


\renewcommand\bibsection{}
\bibliographystyle{thesis_bibtex}

\bibliography{mybibliography}

\end{document}